\documentclass[graybox]{svmult}

\usepackage{mathptmx}       % selects Times Roman as basic font
\usepackage{helvet}         % selects Helvetica as sans-serif font
\usepackage{courier}        % selects Courier as typewriter font
\usepackage{type1cm}        % activate if the above 3 fonts are
\usepackage{makeidx}         % allows index generation
\usepackage{graphicx}        % standard LaTeX graphics tool
\usepackage{multicol}        % used for the two-column index
\usepackage[bottom]{footmisc}% places footnotes at page bottom
\usepackage{natbib}

\makeindex             % used for the subject index
\begin{document}

\title*{Modeling magnetospheric fields in the Jupiter
  system}
%\title*{Interactions of planetary magnetic fields in the Jupiter
 % system}
\titlerunning{Magnetospheric fields in  Jupiter system} % for an abbreviated version of
\author{Joachim Saur, Emmanuel Chan{\'e} and Oliver Hartkorn}
\institute{Joachim Saur, Institute of Geophysics and Meteorology, 
  University of Cologne, Albertus Magnus Platz, 50937 Cologne, Germany
  \email{saur@geo.uni-koeln.de},
\and  Emmanuel Chan\'e \at Centre for mathematical Plasma Astrophysics, KU
Leuven, Celestijnenlaan 200B, 3001 Leuven, Belgium
\email{emmanuel.chane@kuleuven.be},
\and Oliver Hartkorn \at Institute of Geophysics and Meteorology, 
  University of Cologne, Albertus Magnus Platz, 50937 Cologne,
  Germany,
\email{hartkorn@geo.uni-koeln.de}
}
%\and Name of Third Author \at Name, Address of Institute \email{name@email.address}}
\maketitle

\abstract{
The various processes which generate magnetic fields
within the Jupiter system are exemplary for a large class of similar processes occurring
at other planets in the solar system, but also around extrasolar
planets. Jupiter's large internal dynamo magnetic field 
generates a gigantic magnetosphere, which in contrast to Earth's
magnetosphere is strongly rotational driven and possesses large plasma
sources located deeply within the magnetosphere. The combination of
the latter two effects is the primary reason for Jupiter's main
auroral ovals. Jupiter's moon
Ganymede is the only known moon with an intrinsic dynamo magnetic
field, which generates a mini-magnetosphere located within Jupiter's larger
magnetosphere including two auroral ovals. Ganymede's mini-magnetosphere is qualitatively
different compared the one from Jupiter. It possesses no bow shock but
develops pronounced Alfv\'en wings similar to most of the extrasolar planets
which orbit their host stars within 0.1 AU. New numerical models of
Jupiter's and Ganymede's magnetospheres presented here
provide quantitative insight into these magnetospheres and the
processes which maintain them. Jupiter's
magnetospheric field is time-variable on various scales. At the locations of Jupiter's
moons time-periodic magnetic fields induce secondary magnetic fields
in electrically conductive
layers such as subsurface oceans. In the case of Ganymede, these secondary magnetic fields
influence the oscillation of the location of its auroral ovals. Based on dedicated
Hubble Space Telescope observations, an analysis of the amplitudes of the auroral
oscillations provides evidence that Ganymede harbors a subsurface
ocean. Callisto in contrast does not possess a mini-magnetosphere, but
still shows a perturbed magnetic field environment generated by
induction within an electrically conductive layer and due to the plasma interactions
with its atmosphere. Callisto's ionosphere and atmospheric UV emission is
different compared to the other Galilean satellites as it is
primarily been generated by solar photons compared to
magnetospheric electrons. At Callisto a fluid-kinetic model of the ionospheric electron
distribution provides constraints on Callisto's oxygen atmosphere.
}
\section{Introduction}
\label{sec:1}

The plasma interactions in and around Jupiter's magnetosphere and
around Jupiter's moons are so
rich in various phenomena that a huge class of interactions occurring
at other planetary bodies
in the solar system and at extrasolar planets are
represented by the processes in the Jupiter system. In this chapter,
we will present results of new models 
of the interaction of
Jupiter's magnetosphere with the solar wind and interactions at the
Galilean moons with particular focus on Ganymede and Callisto.

Jupiter is the largest planet with the largest magnetic
moment in the solar system. It's interaction with the solar wind
generates a planetary magnetosphere, which would appear to an observer on
Earth larger compared to how the sun appears to us in the sky
if Jupiter's magnetosphere could be seen with the naked eye.
The four large Galilean moons, Io,
Europa, Ganymede and Callisto are all located within Jupiter's gigantic
magnetosphere, and are subject to the interactions generated by
Jupiter's magnetospheric plasma. The interaction at these moons differ due to
the different properties of the moons. For example, Ganymede possesses
an internal dynamo magnetic field which leads to its own mini-magnetosphere
within Jupiter's magnetosphere. The other moons in contrast only possess
weak internal magnetic fields generated by electromagnetic induction in
electrically conductive layers.

In the next section \ref{s_characterize} we categorize the plasma interactions in
Jupiter's magnetosphere based on various parameters. With this
background we then describe the interaction of the solar wind with
Jupiter in order to better understand Jupiter's magnetosphere (see section \ref{s_Jupiter}). In
the subsequent section  \ref{s_time} we
investigate time-variable effects in Jupiter's magnetosphere, which
can be subdivided in periodic and non-periodic variabilities. The
associated time-variable magnetic fields induce secondary magnetic
fields in electrically
conductive layers within the moons. Measurements of these
induced fields are diagnostic of internal layers, such as saline
subsurface oceans. The plasma of Jupiter's magnetosphere interacts
with the atmospheres and ionospheres and the interior of the
moons. Properties of this interaction, the formation of the
ionospheres and the generation of related magnetic fields will be
discussed for Ganymede and Callisto in sections \ref{s_Ganymede} and
\ref{s_Callisto}. A novel technique to search for induced magnetic
fields from a subsurface ocean within Ganymede based on Hubble Space
Telescope observations  of its
auroral ovals is described in section \ref{s_ocean}.

\section{Characterization and description of the interaction}
\label{s_characterize}
The plasma interactions of flows past planetary bodies can be
characterized into different classes,
which we discuss in the next subsection. Afterwards we provide an
overview of the magnetohydrodynamic (MHD) approach to describe these interactions.

\subsection{Overview of the interaction}
\label{ss_}

The interaction of a planetary body with its surrounding plasma is
controlled by two factors: (1) the properties of the plasma flowing
past the planetary object (discussed in section \ref{sss_mach}) 
and (2) the properties of the planetary body
itself (discussed in section \ref{sss_body}). Other overviews of the
plasma interaction at Jupiter and its moons can be found, e.g., in 
\cite{neub98,kive04} or \cite{krup04}.

\subsubsection{Mach numbers and nature of interaction}
\label{sss_mach}

A key property which controls
the interaction is the ratio of the  relative bulk flow velocity $v_0$
between the plasma and the planetary object compared to the group
velocities of the three magnetohydrodynamic waves. 
These ratios are called the fast Mach number $M_f$, the
Alfv\'en Mach number $M_A$, and the
slow mode Mach number $M_s$ and refer to the ratios of the flow
velocity
to the  the fast magneto-sonic mode, the
Alfv\'en mode and the slow magneto-sonic mode, respectively 
\cite[e.g.,][]{baum96}. If the fast
mode is larger than 1, a bow shock forms ahead of the object
for nearly all object classes. Exceptions are inert moons,
i.e., without atmosphere and intrinsic magnetic fields. For $M_f< 1$, the bow shock
disappears which is always the case if $M_A < 1$, i.e., if the flow is
sub-Alfv\'enic.

The interaction of all planets in the solar system with the solar wind
is such that under average conditions the solar wind flow is
super-fast, i.e., $M_f > 1$ and all planets are surrounded by a bow
shock. Only exceptionally, i.e., approximately once every two years,
the plasma density in the solar wind is so low that the resultant
Alfv\'en velocity $v_A= B/ \sqrt{\mu_0 \rho}$ is faster than the solar wind
velocity with the mass density $\rho$ and the magnetic permeability of
free space $\mu_0$. In this case, the Earth looses its bow shock \cite[]{chan12,chan15}. This
transition has however never been observed for Jupiter.
In the case of Jupiter's moons, which are embedded within Jupiter's
magnetosphere, the relative flow velocity is smaller than the Alfv\'en
velocity and thus no bow shock forms, but so called Alfv\'en wings
develop \citep{neub80,goer80,sout80}. Alfv\'en wings are standing
Alfv\'en waves in the restframe of the moons. The Alfv\'en waves are
generated because the obstacle slows the flow of the magnetized plasma
in its vicinity and generates stresses in the magnetic field. 
The Alfv\'en wings can be described as 'tubes' in direction parallel and
anti-parallel to the magnetic field but with additional tilts by an angle
$\Theta_A \approx \tan^{-1 }M_A$ with respect to the magnetic field
\cite[]{neub80}.  

\begin{figure}[t]
\sidecaption
\includegraphics[scale=.85]{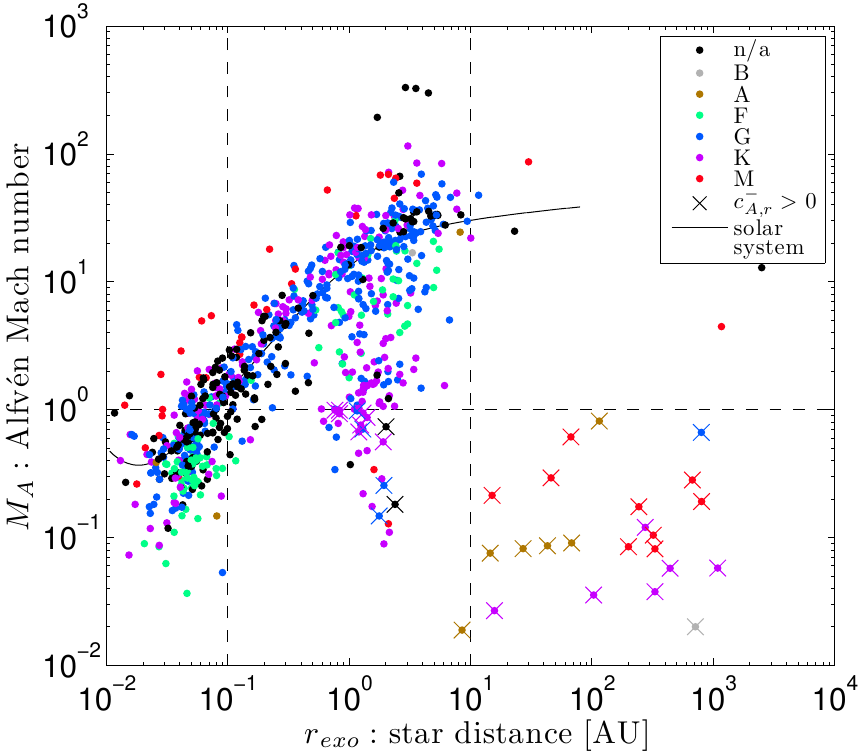}
\caption{Estimated Alfv\'en Mach numbers $M_A$ near all 850 extrasolar planets
  known until 2013. For all extrasolar planets with $M_A < 1$ no bow
  shock forms, which is the case for most of the planets located
  within 0.1 AU of their host star
  \cite[from ][]{saur13}.}
\label{f_exo}   
\end{figure}

Extrasolar planets are observed to orbit their host stars
at a wide range of radial distances from approx 0.01 AU to 1000
AU. Because the stellar plasma properties are expected to strongly evolve as a
function of distance similar to the solar wind properties of the sun
 \cite[e.g.,][]{park58,preu06,lanz08}, the
plasma conditions around the observed extrasolar planets
are expected to vary strongly. In Figure \ref{f_exo}, 
we show the expected Alfv\'en
Mach number $M_A$ near the 850 extrasolar planets known until
2013  \cite[from][]{saur13}. The Mach number is calculated based on measured
and estimated properties of the host stars and by applying the
\cite{park58} model for the radial evolution of the stellar
winds. Figure \ref{f_exo} shows that extrasolar planets at orbital
distances approximately less than 0.1 AU are typically subject to
sub-Alfv\'enic conditions. At these bodies no bows shock but Alfv\'en
wings form. In case such an extrasolar planet possesses a dynamo
magnetic fields as expected \cite[]{chri09}, then the solar system
analogue of these extrasolar planet is Ganymede. In case they do not
possess a dynamo field, their interaction is qualitatively similar to
Io, Europa, and Callisto. 
%{\bf 
This comparison only relates to the sub-Alfv\'enic character of the
  interaction and the principle nature of the obstacle. The atmospheres
of the close-in exoplanets however can be significantly different
compared to those of the Galilean satellites. 
The interaction with
these atmospheres and the planetary magnetic field is expected to 
generate plasma and magnetic field
perturbations qualitatively similar to the moons of Jupiter.
If the exoplanets possess
electrically conductive layers, any time-variable external magnetic
field will additionally induce secondary magnetic fields similar to the
mechanisms at the  Galilean satellites.
%}
Extrasolar planets at stellar separation larger
than approximately 0.1 AU are on average subject to super-Alfv\'enic
conditions ($M_A>1)$ and thus are expected to possess bow shocks 
if additionally $M_f >1$ similar to Jupiter. 
Thus the interactions in the Jupiter system are text book cases for
the interactions at extrasolar planets with the
benefit that a huge set of in-situ and remote-sensing observations are
available. Thus an improvement of
our understanding of the plasma and magnetic field environments in the
Jupiter system is helpful as this understanding is also relevant for 
extrasolar planets.

\subsubsection{The planetary body: Nature of the obstacle}
\label{sss_body}

The nature of the planetary body, i.e., the obstacle to the flow,
shapes the plasma and magnetic field environment in particular very
close to the planetary body. The body can be a mechanical obstacle,
e.g., due to collisions of the neutral particles in the atmosphere with the plasma or if the
body possesses a solid surface. The body can also be an
electromagnetic obstacle, e.g., if it possesses an internal magnetic
field or if it is electrically conductive. The Galilean satellites
are a mix of both mechanical and electromagnetic obstacles. They all
possess very dilute atmospheres and solid, plasma absorbing
surfaces. Among the Galilean satellites, Ganymede is the strongest
electromagnetic obstacle because it possesses an internal dynamo
magnetic field, which generates a mini-magnetosphere
within Jupiter's large magnetosphere \citep{kive96b}. But electrically
conductive layers within all of the moons, e.g. saline oceans,
metallic cores or a possible magma ocean, generate induced magnetic
fields, which influence the external space and plasma environment
around all of theses moons \cite[e.g.,][]{khur98,neub98,khur11,seuf11}.

\subsection{MHD model}
\label{ss_MHD}

The most common approach to describe the interaction of a planetary
body within its surrounding plasma is the MHD approach. It describes
the plasma as an electrically conductive fluid and is applicable to
describe the overall properties of the interaction if typical length
scales of the interaction are larger than the ion gyro radius and
typical time scales are larger than the ion gyro period. 

The MHD approach applies to describe the temporal and
spatial evolution of the mass density
$\rho_m$ the continuity equation, for the plasma bulk velocity
$\vec{v}$ the velocity equation, for the magnetic field $\vec{B}$ the induction equation,
and an equation for the internal energy $\epsilon$, respectively, 
\begin{eqnarray}
\partial_{t}\rho_m+\nabla\cdot(\rho_m \vec{v})&=&(P-L)m_i \; ,\label{e_konti} \\
\rho_m\partial_{t}\vec{v}+\rho_m\vec{v}\cdot\nabla\vec{v}&=&
 -\nabla
                                                             p+\left(\frac{1}{\mu_0}\nabla\times\vec{B}\right)\times\vec{B}\,
                                                             {-(\rho_m\nu_{in}
                                                             +Pm_i)
 \vec{v} 
% (\vec{v} -\vec{v_n} )
} \; , \label{e_momentum}   \\
\partial_{t}\vec{B} &=&\nabla\times\left(\vec{v}\times\vec{B}\right)
                        + \eta \Delta \vec{B} \;, \label{e_induction} \\
\partial_t\epsilon+\nabla\cdot(\epsilon\vec{v})&=&-p\nabla\cdot\vec{v}
+\frac{1}{2} \rho_m \vec{v}^2\left(\frac{P}{n}+\nu_{in}\right)
-\epsilon \left(\frac{L}{n}+\nu_{in}\right)\label{e_energy} \; .
\end{eqnarray}
The continuity equation (\ref{e_konti}) can include sources $P$
due to ionization of neutral particles with mass $m_i$ and losses $L$ due to
recombination. The velocity equation (\ref{e_momentum}) includes in
case of a mechanical obstacle the collisions and the mass loading in
an atmosphere of the planetary body with $\nu_{in}$ the collision
frequency for momentum transfer between the ions and the neutrals
through elastic collisions and charge exchange. 
Here we assumed that the velocity of
the atmosphere is at rest.  The evolution of the
magnetic field is described by the induction equation
(\ref{e_induction}), which can include a resistive term characterized
by the magnetic diffusivity $\eta$. The evolution of the internal energy density
$\epsilon$, which is related to the plasma thermal pressure $p$
through $\epsilon=3/2\, p$ is described through (\ref{e_energy}).
The total thermal pressure $p$ includes the effects of the electron temperature $T_e$
and the ion temperature $T_i$   by
$p=nk_B(T_e+T_i)$  with the Boltzmann constant $k_B$, and the plasma number
density $n$. The MHD approach does however not allow to constrain the
$T_e$ and $T_i$ separately without further assumptions.
 In (\ref{e_energy}),
next to the work done by the pressure, the collisions of the plasma
with neutrals as well as plasma production and recombination are
included.

In order to describe the interaction as a well posed problem, initial and boundary
conditions need to be specified. For the initial conditions, the
unperturbed plasma or approximations to the final solutions are
generally used. The outer boundary conditions are chosen so
that on the upstream side of the obstacle inflowing conditions are
set. They characterize the properties of the plasma flow upstream of the
obstacle. Downstream of the obstacle outflowing boundary conditions
are applied \cite[e.g.,][]{chan13,duli14}. The inner boundary for the
plasma is located at the surface of the planetary body if the plasma
reaches all the way to the surface \cite[]{duli14}. For a planet with a very dense
atmosphere such as Jupiter, the inner boundary is located below the
ionosphere \cite[]{chan13}. In case of the moons, the plasma is absorbed at
the surfaces of the solid bodies, which are additionally assumed to be no source of
plasma. These conditions imply that the radial component of the
plasma flow $v_r$ can only be negative or zero, which also
sets the conditions for the plasma, momentum and energy flow in
equations (\ref{e_konti}), (\ref{e_momentum}), and
(\ref{e_energy}) \cite[]{duli14}. In case of Jupiter the inner boundary conditions for
the plasma is such that $v_r=0$ \cite[]{chan13}. The boundary
condition for the magnetic field is given by the electrically
non-conductive nature of the solid surface of the moons and the
neutral atmosphere below Jupiter's ionosphere. This insulating nature implies
that the radial component of the electric current at the boundary
needs to vanish, i.e., $j_r=0$. The latter condition has non-local
effects on the magnetic field and can be implemented by decomposing the
magnetic field at the surface into poloidal and toroidal fields,
which are expended into spherical harmonics. \cite{duli14} showed that for the
resulting complex poloidal and toroidal coefficients $p^*_{lm}$ and
$t^*_{lm}$ of the  spherical harmonics $Y^*_{lm}$ of
degree $l$ and order $m$, respectively,
the following two equations need to be fulfilled
\begin{equation}
 t^*_{lm}(R_0,t)=0, \label{tconanal}
\end{equation}
\begin{equation}
 R_0\frac{\partial p^*_{lm}(r,t)}{\partial r}\bigg|_{r=R_0}- (l+1) p^*_{lm}(R_0,t) =-\frac{2l+1}{l}G^*_{lm}(t). \label{pconanal}
\end{equation}
These equations need to be fulfilled at the inner boundary at every
time step and thus set the magnetic field boundary
conditions at the surface located at $R_0$. This description of the
inner boundary conditions for the magnetic field also allows that the
planetary body possesses internal magnetic fields whose origin lie below
the surface. They can be internal dynamo fields as is the case for
Jupiter and Ganymede, but they can also be time-variable induction
magnetic fields generated in saline electrically conductive subsurface
oceans. The internal fields are represented in (\ref{pconanal}) through
their complex Gauss coefficients $G^*_{lm}(t)$. 
In case of Jupiter and its very strong magnetic field, the
inner boundary conditions can be approximated by setting the azimuthal
and the longitudinal component of the magnetic field to the internal
Jovian magnetic field \cite[]{chan13}.

%{\bf 
The MHD approach is overall a very powerful approach to describe
  the plasma dynamics in the Jupiter system. Naturally, it 
does not capture all aspects of the plasma interaction, e.g., if electron
scale physics is relevant and/or gyro 
radii and  gyro periods need to be resolved. 
For example, reconnection at the  magnetopauses of Jupiter and
Ganymede is an non-ideal MHD effect which is best described with models
which resolve ion and electron kinetics, e.g., with full particle-in-cell
models.  Finite gyro radii effects can play a role in situations where
non-thermal, high energy ions are involved or where ions with large
velocities within a small background magnetic field are
picked-up. In case the latter effects are important, appropriate
models are hybrid models or particle-in-cell models.  Even though
MHD models cannot resolve ion and electron kinetics effects, one of
their advantages in numerical simulations is that they generally enable better
spatial resolutions compared to  kinetic models in case of similar
computational resources. This advantage is particular helpful if
small atmospheric scale heights need to be resolved.
%}

\section{Jupiter's magnetosphere}
\label{s_Jupiter}
Jupiter's magnetosphere is qualitatively different compared to the
magnetosphere of the Earth. 
The magnetosphere of Jupiter is
rotationally dominated throughout a large part of the magnetosphere 
and the moons of Jupiter are huge internal mass
sources \cite[e.g.,][]{vasy83,bage11}. Io provides approximately
10$^3$ kg s$^{-1}$ mostly in form of SO$_2$ and Europa approximately 50 kg s$^{-1}$
in form of O$_2$ to the magnetosphere
\cite[e.g.,][]{broa79,saur98,saur03a,mauk03}. This mass is subsequently being ionized and
picked up by the motional electric field of the fast rotating
magnetosphere. Jupiter's sideric rotation period is $\sim$9.9 h. 
The plasma thus experiences large centrifugal forces
responsible for radial transport of the plasma. Due to conservations
of angular momentum the magnetospheric plasma while being transported
radially outward is not rigidly corotating anymore. This sets up
magnetic stresses and electric current systems which couple the
magnetosphere to Jupiter's ionosphere. Through this coupling angular
momentum  is being transported from Jupiter's ionosphere into Jupiter's
magnetosphere to bring the magnetosphere closer to full corotation.
Jupiter's magnetosphere and its internal coupling 
has been described theoretically and numerically by a
series of authors 
\cite[e.g.][]{hill79,vasy83,hill01,cowl01,sout01,walk98,walk03,mori08,ray10}.

Because of the immense importance of Jupiter's
magnetosphere-ionosphere (MI) coupling, \cite{chan13} developed a new
MHD model of Jupiter's magnetosphere. This model explicitly includes
the ionosphere of Jupiter within the model domain. In Jupiter's ionosphere
ion-neutral collisions transfer angular momentum from the neutrals
onto the plasma. The resultant flow generates magnetic stresses
between the ionosphere and magnetosphere, which accelerate the
magnetosphere in the direction of Jupiter's rotation, but decelerate
Jupiter's ionosphere in return. In the model of \cite{chan13} the
MI-coupling is explicitly included and the magnetic field boundary
conditions can be physically correctly set below the ionosphere (see
section \ref{ss_MHD}). The downside of the approach is that for
numerical reasons, the radial extension of the ionosphere is strongly
exaggerated by 4 Jovian radii (R$_J$) and the surface of Jupiter has been
set to be at 4.5 R$_J$. Despite the latter assumption, the
\cite{chan13} model however still uses the closest inner boundary of
all published MHD models of Jupiter's magnetosphere. The model  
includes the mass loading in the
Io plasma torus explicitly. It thus explicitly includes two of the most
important features of Jupiter's magnetosphere: The coupling to the
ionosphere and the mass loading in a fast rotating magnetosphere.

\begin{figure}[t]
\sidecaption
\includegraphics[scale=.41]{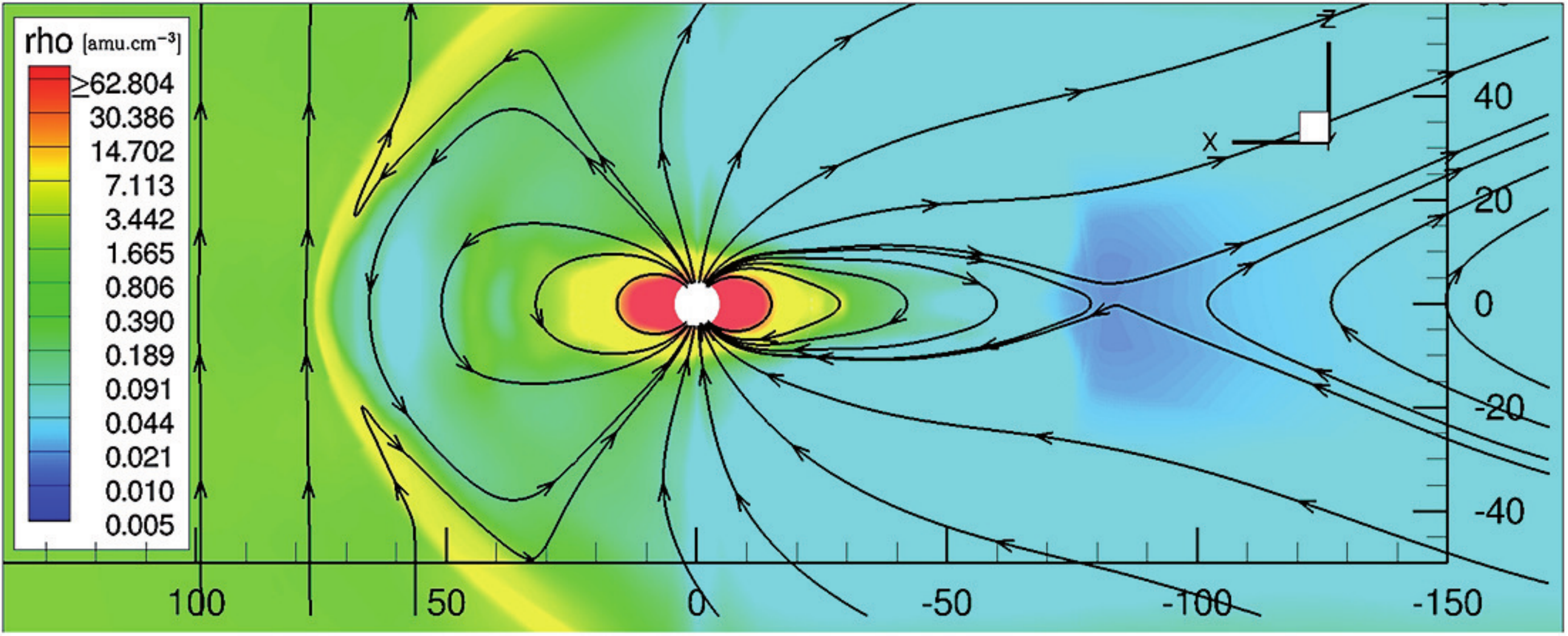}
\caption{Density contours in the noon-midnight meridian of Jupiter's
  magnetosphere. The solar wind is coming from the left. The magnetic field lines are shown in black
  \cite[from ][]{chan13}.}
\label{f_Jupiter_rho}   
\end{figure}

The overall density and magnetic field structure of the modelled
magnetosphere is shown in Figure \ref{f_Jupiter_rho}.
The bow shock is located at 73 R$_J$ and the magnetopause at 69 R$_J$
in agreement with in-situ measurements \cite[]{joy02}. The
magnetosphere is compressed on the day side and strongly elongated on
the night side. Figure \ref{f_Jupiter_rho} also shows an X-point on
the night side where a plasmoid of plasma is being released as part of
the mass loss processes in the tail of the magnetosphere. The figure
also shows that the plasma is concentrated in the equatorial regions
of the magnetosphere due to the strong centrifugal forces. The radial
density profile of the \cite{chan13} model agrees very well with measured density
profiles taken by the Galileo spacecraft \cite[]{fran02,bage11}.

\begin{figure}[t]
\sidecaption
\includegraphics[scale=0.65]{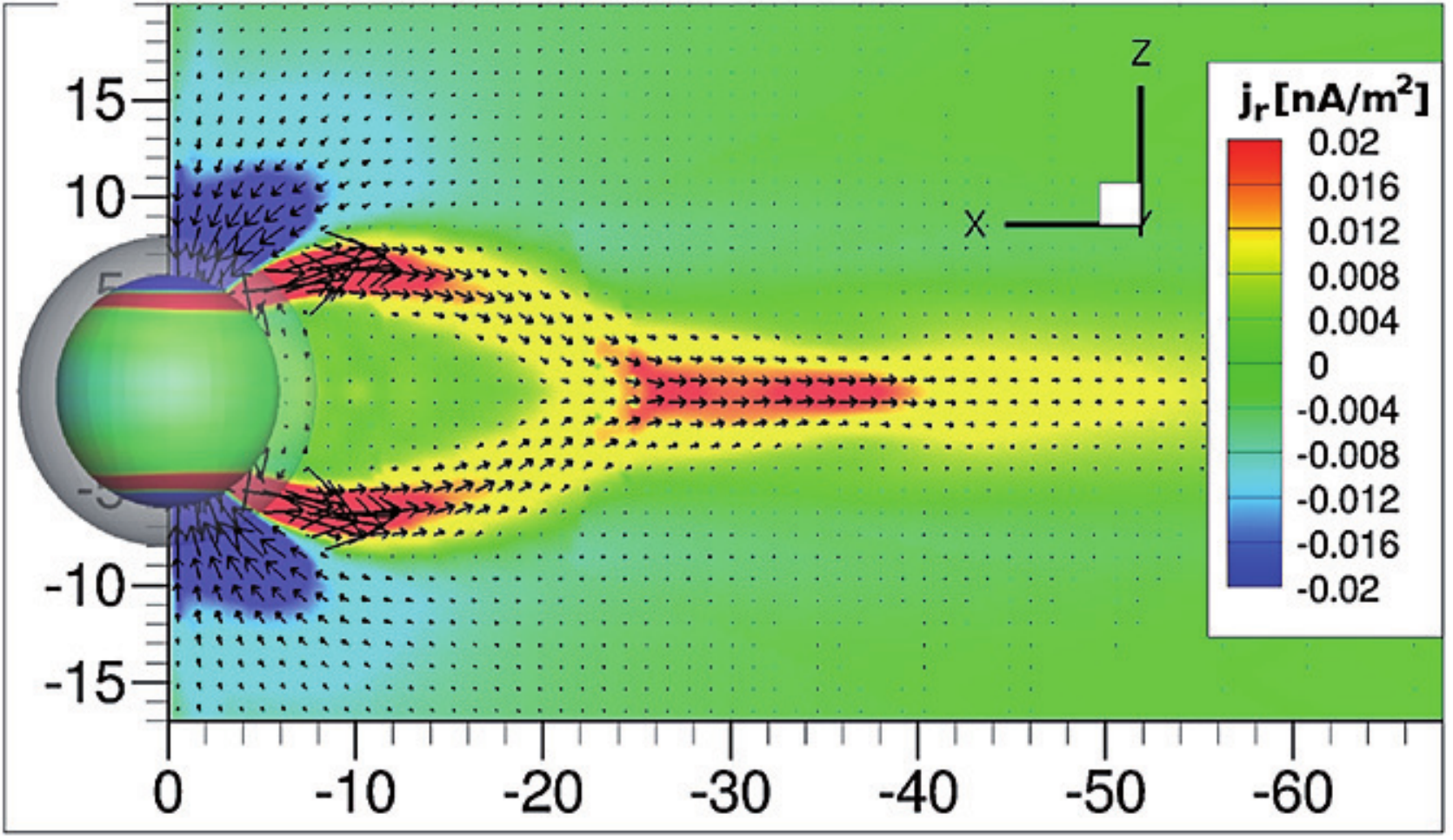}
\caption{
Corotation enforcing current system on the
nightside of Jupiter in the noon-midnight meridian. 
The radial current is shown with color contours,
and the direction of the current in the same plane 
is represented by the black arrows. The radial
current in the ionospheric region is displayed on a
sphere at 6 R$_J$, and a transparent sphere at 8 R$_J$ shows the
extend of the ionospheric region. Note that the corotation
breakdown of rigid corotation
in this plane occurs at 32 R$_J$ 
  \cite[from ][]{chan13}.
 }
\label{f_Jupiter_j}   
\end{figure}

In Figure \ref{f_Jupiter_j},
the electric current system which represents  the angular momentum
transfer between Jupiter's ionosphere and magnetosphere is
displayed. The electric current loop connects Jupiter's ionosphere
through field aligned currents directed away from Jupiter and which
are feed into the equatorial plasma sheet of Jupiter's magnetosphere 
at radial distances of 20 to 30 R$_J$. In this plasma sheet the
currents are directed mostly radially outward and the related $\vec{j}
\times \vec{B}$ forces spin up the magnetosphere. The current closure
back to the ionosphere occurs at large radial distances (not clearly
visible at low latitudes in Figure (\ref{f_Jupiter_j})) and enter the
ionosphere in the polar region. The $\vec{j} \times \vec{B}$ forces
slow the plasma in the ionosphere and are in balance with the forces
exerted by the ion-neutral collisions, which accelerate the ionosphere.
%The magnetospheric velocity profiles are thus strongly controlled by  
%$\vec{j} \times \vec{B}$ forces. 

\begin{figure}[t]
\sidecaption
\includegraphics[scale=0.6]{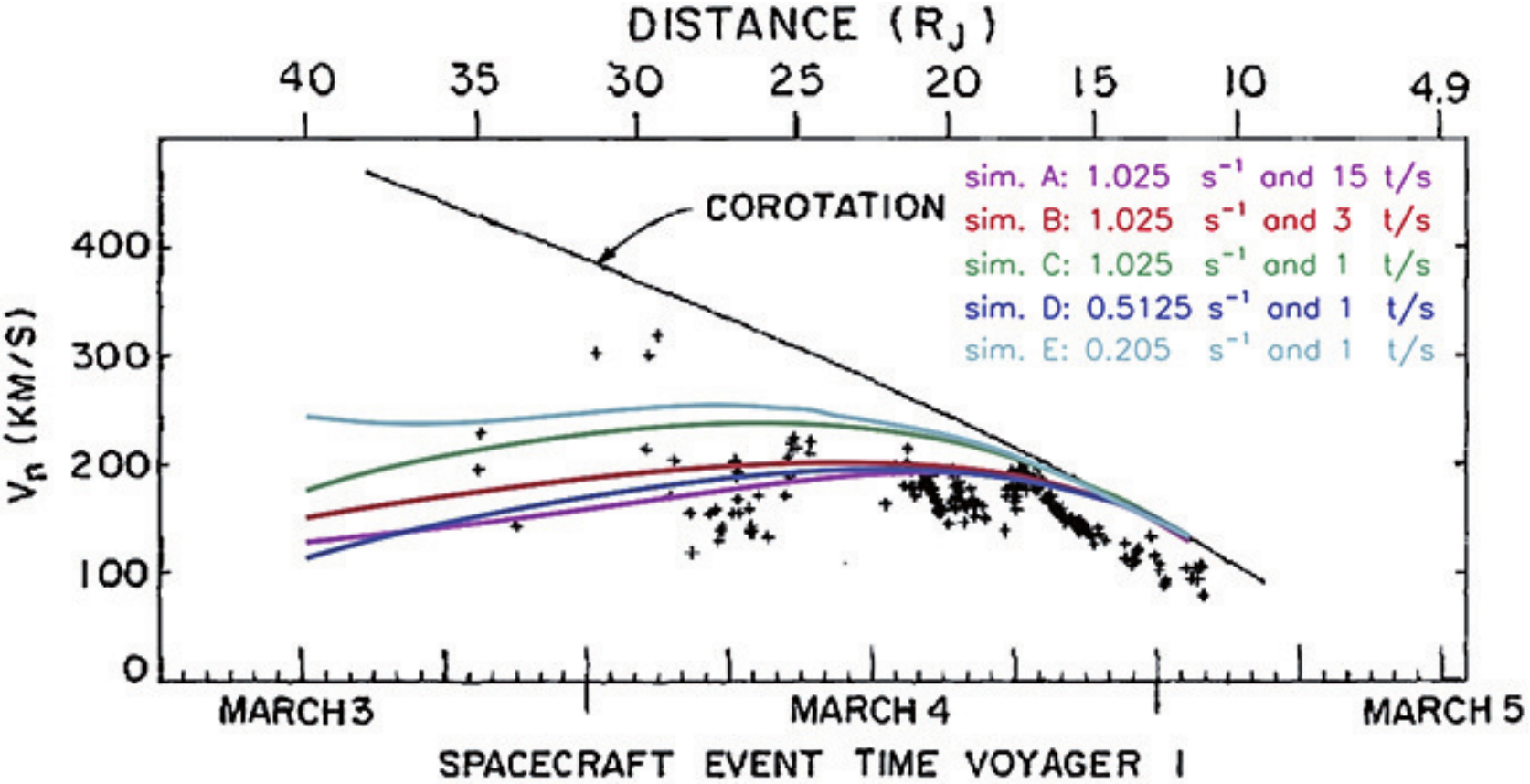}
\caption{
Azimuthal velocity as a function of radial
distance: Comparison between Voyager I measurements
(black plus signs) and computational results (color lines).
The model output values are over plotted on top of
Figure 6.24 from \cite{khur04}, which was adapted
from \cite{mcnu81}. The model values are shown for
different Io torus mass loading rates and ionospheric ion-neutral
collision frequencies. The black line represents rigid
corotation. For the computations, the values given are for the
equatorial plane and were averaged over a rotation period
and over all local times.
  \cite[from ][]{chan13}}
\label{f_Jupiter_velocity}   
\end{figure}

The modeled azimuthal and radial velocities
in \cite{chan13} are in good agreement with observations by the
Galileo and Voyager spacecraft and the theoretical predictions by
\cite{hill79}. In Figure \ref{f_Jupiter_velocity} the modeled azimuthal
velocities as a function of radial distance are shown for various mass loading
rates  in comparison to Voyager measurements.  
 The corotation
breakdown (as defined by \cite{hill79}, 75\% of rigid corotation)
in the plane of Figure \ref{f_Jupiter_j} occurs at  a radial distance
of 32 R$_J$, but the plasma starts
to subcorotate at approximately 20 R$_J$. These results are consistent with the
locations of the field-aligned currents  and matches the
theoretical predictions obtained for the same ionospheric conductances
and total radial mass transport rates by \cite{hill79,hill01}.

\begin{figure}[b]
\sidecaption
\includegraphics[scale=0.55]{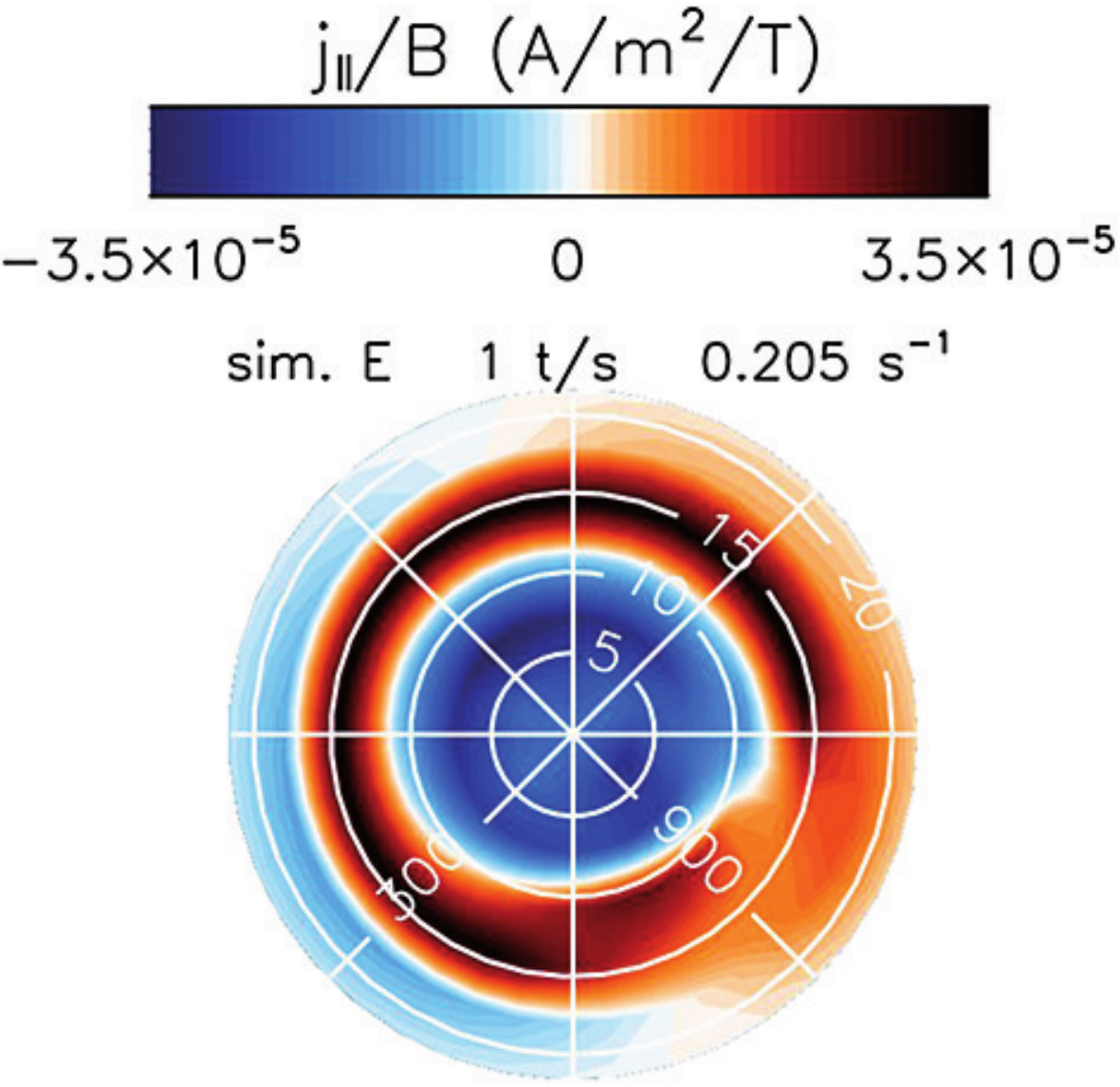}
\caption{
Color contours show $j_\parallel/
B$ in Jupiter's northern hemisphere
at 1 $R_J$. The values of the current were projected from
the ionosphere to a sphere at 1 R$_J$ by following the dipole
magnetic field lines. The value of $j_\parallel/
B$ is averaged over a
rotation period. The colatitude and the local time are plotted
in white on the figures; the dayside is located on the right
  \cite[from ][]{chan13}.
}
\label{f_Jupiter_aurora}   
\end{figure}

On field lines with large parallel electric currents pointing away
from the planet and on locations along these field lines where the charge
carrier density is small, electrons need to be accelerated to large
energies to maintain the electric current loop
\cite[]{knig73,hill01,cowl01}. The resultant energetic electrons
precipitate into Jupiter's ionosphere and excite Jupiter's main
auroral oval, which has been extensively observed, e.g., with the
Hubble Space Telescope (HST) \cite[e.g.,][]{clar05}. Therefore the location in
Jupiter's atmosphere, where field lines with large anti-planetward electric
current densities map to, can be associated with auroral emission. In
Figure \ref{f_Jupiter_aurora}
 we show the electric current density from the \cite{chan13} model
mapped along dipole field lines into Jupiter's atmosphere. The images
show that the current system matches to colatitudes of approximately
15$^\circ$, which is in good agreement with observations
\cite[e.g.,][]{clar05}. Figure \ref{f_Jupiter_aurora} also shows that
the electric current is azimuthally asymmetric as expected from a
magnetosphere with strong local-time asymmetries. In particular,
between 8:00 and 11:00 LT the anti-planetward electric current density
has a minimum, which is consistent with a discontinuity in the
main oval observed by \cite{radi08} within HST observations.
The discontinuity in the electric current in the model of \cite{chan13} 
is caused by an asymmetry in the pressure distribution due
to the interaction between the rotating plasma and the magnetopause.

\section{Time-variable magnetosphere }
\label{s_time}

While we discussed in the previous section mostly steady state components of  
Jupiter's magnetosphere,  we address here magnetospheric
time-variability  caused by the solar wind, by variations of the internal sources, 
by the rotation of Jupiter and by dynamical non-linear processes in the magnetosphere.

The internal plasma sources of Jupiter's magnetosphere, i.e., the mass
loading at the moons might be time-variable which is however observationally not
established very well. The model of \cite{chan13} shows that if
the mass loading rate of Io changes from $1
\times 10^3$ kg s$^{-1}$ to $3
\times 10^3$ kg s$^{-1}$, the azimuthal velocity profile changes and
the break down of corotation occurs further inside (see Figure
\ref{f_Jupiter_velocity}). The change in the mass loading also implies
changes in the size of the magnetosphere, the structure of the magnetic
field, and in the field aligned auroral current systems.
For enhanced mass-loading rates, \cite{chan13} find that the auroral
becomes more symmetric, while the brightness barely changes.
%, which lead to
%enhanced auroral emission for increased mass loading.
%EC's comment: for enhanced mass-loading rates, in the simulations, the main oval becomes more symmetric, but not brighter

Changing solar wind conditions impact Jupiter's magnetosphere as
well. An increase in solar wind ram pressure, decreases the size of
the magnetosphere on the subsolar direction and stretches it towards
the night side. An important question is how the auroral brightness
responds to an increased ram pressure. Earlier theoretical work by \cite{cowl01}, \cite{sout01}
%EC's comment: Cowley and Bunce (PSS, 2001)
and \cite{cowl03b}, suggested that the aurora dims in this case
because the decreased size of the magnetosphere moves the 
magnetospheric plasma somewhat radially inward, leading to 
increased angular velocities (since the angular momentum is conserved). 
This is expected to reduce the
corotational enforcing electric currents and thus to lead to a reduced auroral
brightness. The simulations of \cite{chan13} predict that the
overall response to increased solar wind ram pressure strongly depend on
local time, but lead in general to an overal increased
auroral brightness \cite[]{chan17}. The primary reason is that the
enhanced solar wind ram pressure increases the magnetic stresses in the
magnetosphere leading to higher field-aligned electric currents.
For deriving these results, the three-dimensional nature of
the magnetosphere needs to be considerer. \cite{chan17} find that
only for a short period of time, during the transition phase from weak to strong
solar wind ram pressure the aurora locally dims in the noon region.   

Another cause of time-variability are dynamical processes in the
magnetosphere. Among these processes are intermittent reconnection on
the night side and the release of plasmoids on time scales on the
order of 10 hours 
%(Emmanuel can you check and modify and add other references) 
%EC's comment: we can add Vogt et al. (JGR, 2014) and Bagenal (J. Atmos. Sol. Terr. Phys, 2007)
\cite[]{bage07,chan13,vogt14,chan17}. The radial transport within the
magnetosphere occurs through flux tube interchange
\cite[e.g.][]{kive97}. The resultant perturbations of the
magnetosphere are stochastic in nature. These perturbations in turn
interact with each other generating a turbulent cascade of
time-dependent magnetic field and velocity fluctuations \cite[]{saur02}.

Another class of time-variability is induced in Jupiter's magnetosphere
due to the approximately 10$^\circ$ tilt of Jupiter's magnetic moment
with respect to its spin axis. This tilt makes all properties of
Jupiter's magnetosphere  time-variable with a period of the sideric
rotation period of Jupiter (9.9 hours) when observed in an inertial
rest frame. This time-variability is important for the magnetosphere
itself, but it also leads to time-variable magnetic fields at the
locations of the moons, which can be used to probe their interior
structure \cite[e.g.,][]{khur98,neub98,kive00,zimm00}. 

\begin{figure}[t]
\sidecaption
\includegraphics[scale=1.]{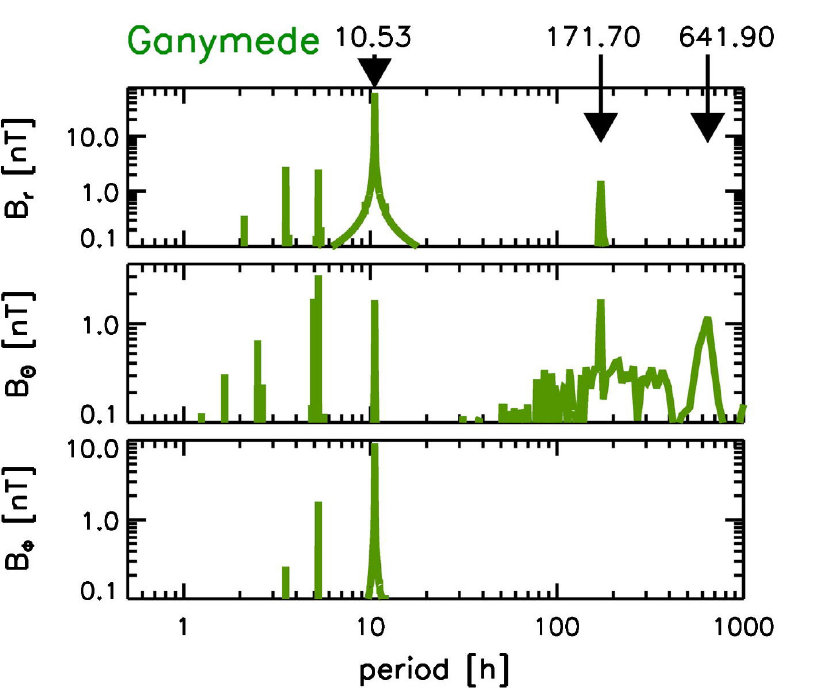}
\caption{Amplitudes of the time-variable radial $B_r$, azimuthal $B_\Phi$ and
  latitudinal $B_\Theta$ magnetic
  field components of Jupiter's magnetosphere at the location of
  Ganymede.  The period
  10.53 h is the synodic rotation period of Jupiter as seen from
  Ganymede, 171.7 h is the orbital period of Ganymede, and 641.9 h is
  solar rotation period 
% Emmanuel, you had changed synodic to chap? I am not familiar with
% the expression.
  \cite[from ][]{seuf11}.
}
\label{f_Ganymede_time_variable}   
\end{figure}

The time-variability of the magnetic field near the Galilean moons  
has been explored extensively by \cite{seuf11} 
through considerations of a range of possibly frequencies. 
The magnetic field model of \cite{seuf11} includes (a) the dynamo
magnetic field of Jupiter represented
by an expansion in spherical harmonics, (b) the magnetic fields of the current sheet 
and (c) fields due to the magnetopause boundary currents. With this
magnetic field model, field components at the location of the moons are
calculated and subsequently Fourier-transformed to obtain the
time-variable magnetic field amplitudes as a function of their
period as shown in Figure \ref{f_Ganymede_time_variable}.
The figure shows three sources of time-variability at Ganymede,
i.e., three sets of periods: (i) the rotation
period of Jupiter and higher harmonics thereof due to non-dipole components
of the interior field and due to non-sinusoidal components of the current sheet
field. The synodic rotation period of Jupiter seen in the restframe of
the satellites generates the strongest amplitudes
of all periods with the maximum in the $B_r$ component of $\sim80$ nT
\cite[]{seuf11}. The higher order harmonics are already significantly
smaller on the order of a few nT or less. (ii) The time-periodic
contribution due to the orbital period of Ganymede is fairly small
on the order of 1-2 nT due the small inclination $i=0.17^\circ$ and
small eccentricity $e=0.0011$ of Ganymede. (iii) The solar rotation period of the
sun is propagated out through the solar wind and can generate
time-variable solar wind ram pressure, which generates time-variable
magnetopause currents also called Chapman-Ferraro currents. The
related currents cause amplitudes in the latitudinal $B_\Theta$ components of less
than 1 nT. We will see in Section \ref{s_Ganymede} that these various
time-dependent fields
generate induced magnetic fields in the interior of
Ganymede. Observations of these induced magnetic field with the
knowledge of the time-variable inducing fields from Figure
\ref{f_Ganymede_time_variable} or previous studies, e.g., by
\cite{kive02} can be used to probe the interior of Ganymede and the
other moons. \cite{seuf11} calculate the amplitudes and the phases 
of the induced magnetic fields for the three inducing frequencies and
for various models of the electrical conductivity
structure within Ganymede, e.g., with a poorly conductive surface, a
conductive saline subsurface water ocean and very highly conductive
metallic core. 
%(HERE WE COULD ADD FIGURE 11 or 12 of SEUFERT ET
%AL. 2011).

\section{Ganymede's magnetosphere}
\label{s_Ganymede}
Ganymede is the largest moon in the solar system and comparable in
size to Mercury. It is also the only known moon with an intrinsic
dynamo magnetic field. Thus it possesses a mini-magnetosphere within
Jupiter's gigantic magnetosphere as studied by a number of
authors, e.g., \cite{kive98,kive02,kopp02,ip02,paty04,paty06,paty08,jia08,jia09,jia10a}. 

\begin{figure}[t]
\sidecaption
\includegraphics[scale=0.44]{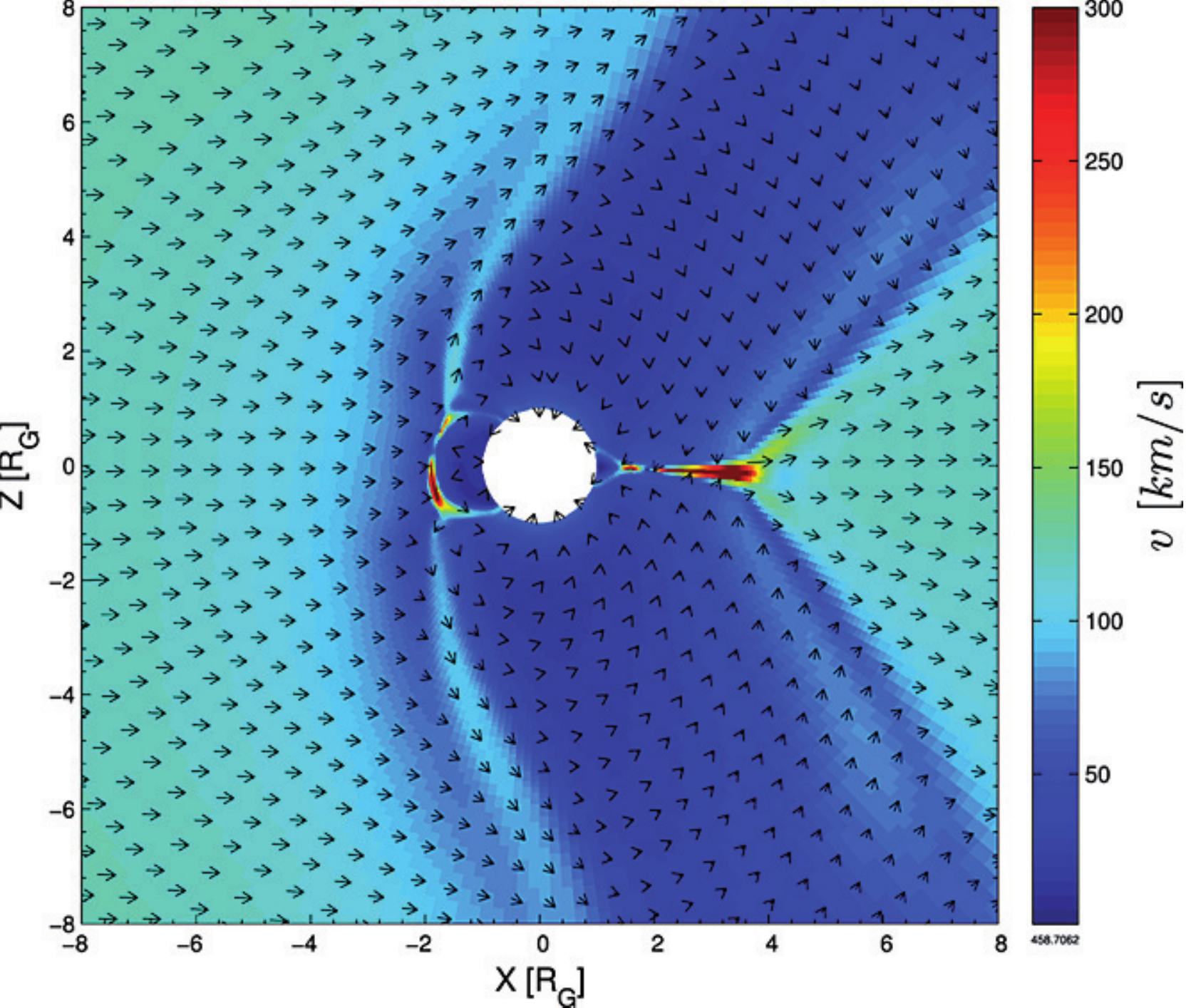}
\caption{Plasma velocity $\vec{v}$ around Ganymede  in a plane
  perpendicular to the direction Ganymede-Jupiter, i.e., the plane given by
  the unperturbed plasma flow and the north-south direction (for the conditions
  of Galileo   spacecraft G8 flyby). 
The length of an arrow in this figure represents the magnitude of
the vector components within the displayed
plane. The total magnitude of the vector is displayed color coded.
%
%The arrows show the tangential parts of the field, and their lengths
%linearly scale with 
%the highest tangential magnitude of this plane displayed as small number below the
%color bar. 
In this figure the color bar is caped while maximum values of
$\sim$460 km s$^{-1}$ are reached in the dark red regions 
  \cite[from ][]{duli14}.
}
\label{f_Ganymede_wings}   
\end{figure}

With the new MHD model developed for
Ganymede by \cite{duli14} and introduced in section \ref{ss_MHD}, we
model Ganymede's plasma and magnetic field environment as displayed in
Figure \ref{f_Ganymede_wings}.
The MHD model includes Ganymede's
internal dynamo magnetic field after \cite{kive02} and
induction in a subsurface ocean through the non-conducting boundary
conditions given in equations (\ref{tconanal}) and
(\ref{pconanal}). The model uses appropriate outer boundary and initial
conditions given by the Galileo spacecraft measurements during each of
its flybys at Ganymede. The model also includes a thin atmosphere, in which
ionization with a constant ionization frequency generates an
ionosphere as a source term in equation (\ref{e_konti}). Elastic
collisions, charge exchange and ionization slow the flow in the ionosphere, which
is implemented in the source terms in the velocity equation (\ref{e_momentum}). In
the induction equation (\ref{e_induction}), we include the resistivity of the ionosphere
and anomalous resistivity due to reconnection similar to \cite{jia10a}.

\begin{figure}[t]
\sidecaption
\includegraphics[scale=0.6]{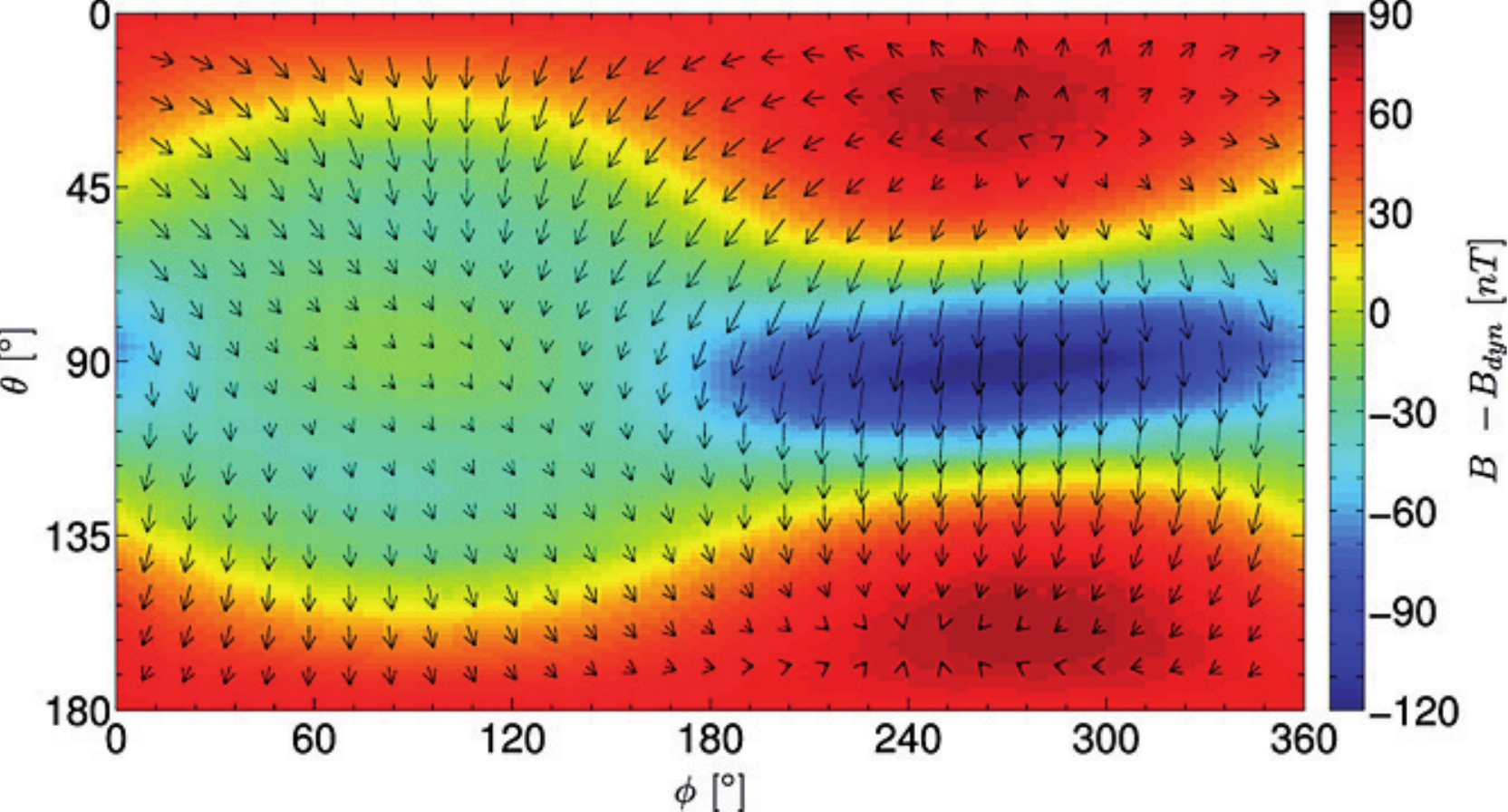}
\caption{Magnetic field component's caused by the
plasma interactions at Ganymede’s surface. Color coded is the difference
of the magnitude of the simulated total magnetic field and the magnitude
of Ganymede’s dynamo-generated magnetic field that is given by
the dipole Gauss coefficients in \cite{kive02}. The arrows show the
tangential components of the plasma magnetic field
  \cite[from ][]{duli14}.
}
\label{f_Ganymede_bc}   
\end{figure}

An appropriate description of the magnetic boundary condition at the
surface of Ganymede (see section \ref{ss_MHD}) is crucial to correctly
describe Ganymede's magnetic field environment. Commonly applied
incorrect boundary conditions are to set the magnetic field at the
surface of Ganymede to fixed values given by the internal magnetic
field. This approach forces the plasma magnetic fields to be zero and
allows electric current to enter through the electrically
non-conducting surface of Ganymede. In Figure
\ref{f_Ganymede_bc},
we show the magnetic field perturbations at the surface of
Ganymede, which would be neglected if the magnetic field at surface is
set to the values of the dynamo field. 
The plasma magnetic field assumes values up to $\sim$120
nT. These values are larger than the time-variable components and the
induction effects of an ocean and are about 20\% of Ganymede's dynamo
magnetic field. Thus applying incorrect boundary conditions
significantly distorts the magnetic field environment around Ganymede. 

\begin{figure}[t]
\sidecaption
\includegraphics[scale=0.35]{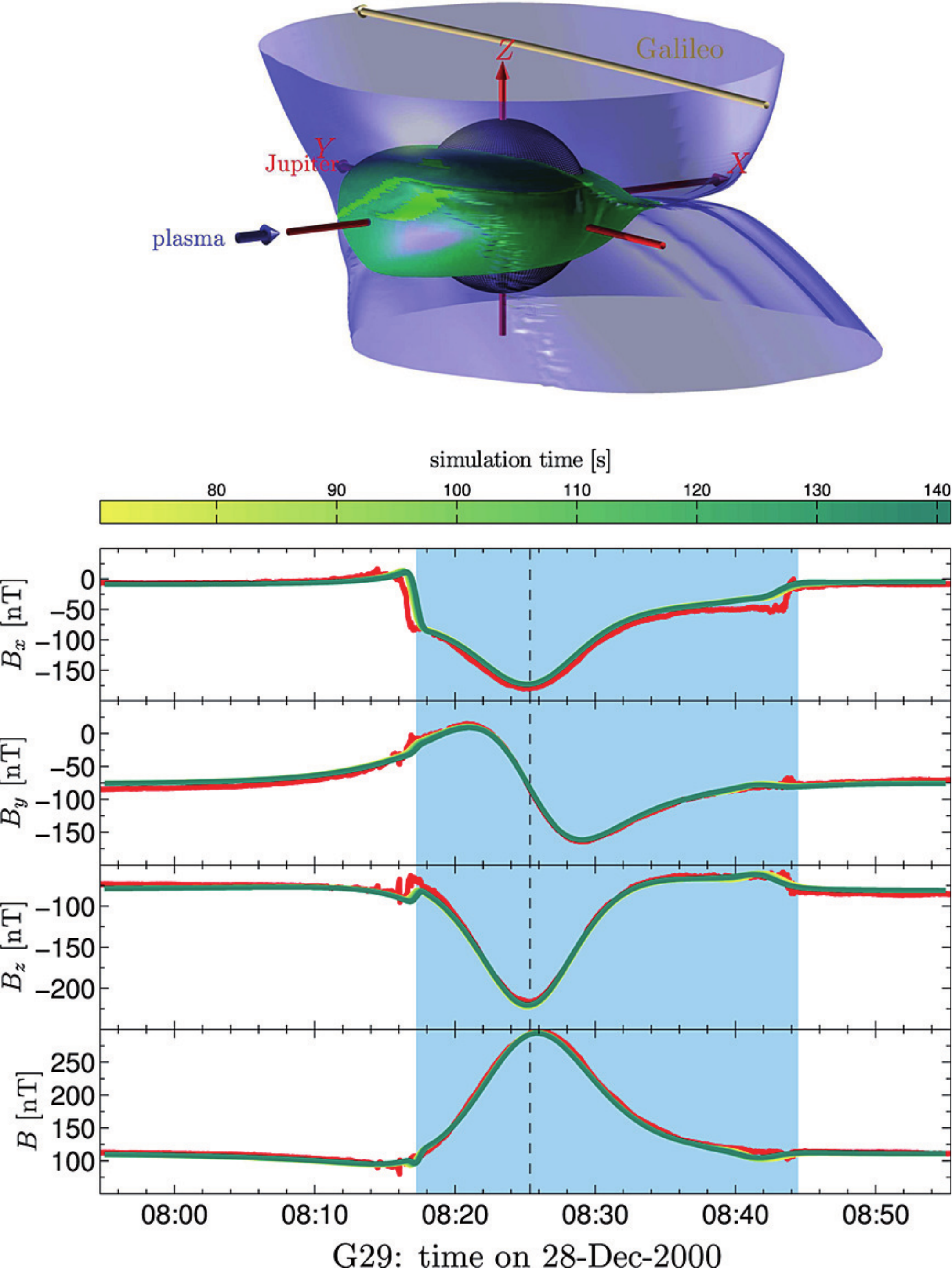}
\caption{Modeled structure of Ganymede's 
magnetosphere in top panel with closed field line region in green and
open field line region in blue. Magnetic field measurements 
in red and model results in yellow-green lines are shown in
bottom. Both panels are for
the G 29 flyby 
  \cite[from ][]{duli14}.
}
\label{f_Ganymede_wing_cross}   
\end{figure}

The relative velocity of Jupiter's magnetospheric plasma with
respect to Ganymede is sub-Alfv\'enic \cite[e.g.,][]{neub98} and thus no bow
shock forms. Ganymede's internal magnetic field generates a
mini-magnetosphere with a region of closed magnetic field lines as can
be seen as the green shaded region in the top panel of Figure
\ref{f_Ganymede_wing_cross}.  
The closed field line region is shifted
towards higher latitudes on the upstream side and is shifted towards
the equator on the downstream side due to the magnetic stresses acting
on Ganymede \cite[]{neub98}. This effect is also well visible in
Figure \ref{f_Ganymede_wings}. The same figure additionally shows that the
plasma flow within the closed field
line region is reversed, i.e., in the upstream direction compared to
the unperturbed magnetospheric flow.

%*

Due to the sub-Alfv\'enic nature of the incoming flow Alfv\'en wings form, which are
displayed as blue region in the top panel of Figure
\ref{f_Ganymede_wing_cross}. The Alfv\'en wings are also visible in
Figure \ref{f_Ganymede_wings} as the large structures north and south
of Ganymede where the plasma flow is strongly reduced. The width of
the Alfv\'en wings is significantly larger compared to Ganymede and the width
of the closed field region. This is due to the orientation of the
internal magnetic field and the external field of Jupiter's
magnetosphere. The width of the wings at the presence of an internal
magnetic field shown in Fig. \ref{f_Ganymede_wing_cross} is in
 agreement with quantitative
expressions for the width derived in \cite{neub98} and \cite{saur13}.

In the bottom panel of Figure \ref{f_Ganymede_wing_cross} we quantitatively
compare the MHD model results of \cite{duli14} shown in green with magnetic field
measurements by the Galileo spacecraft taken during the G 29
flyby shown in red. This flyby crossed the northern Alfv\'en wings and is displayed
as a yellow arrow in the top panel of Figure
\ref{f_Ganymede_wing_cross}. The MHD model fits the amplitude and
locations of the wing crossing well. It also quantitatively reproduces
the magnetic field measurements of all the other Ganymede flyby by the
Galileo spacecraft \cite[e.g.,][]{duli14}.

\section{Ganymede's ocean}
\label{s_ocean}

The time-periodic magnetic fields in Jupiter's magnetosphere as
discussed in section \ref{s_time} can be used to explore electrically
conductive layers within the moons of Jupiter
\cite[e.g.,][]{khur98,neub98,zimm00}. These fields establish one of the few
currently available methods to search for saline and thus electrically conductive
subsurface oceans. The method is based on the fact that water in its
solid form possesses an electrically conductivity at least 4 orders of
magnitude smaller compared to liquid water with salinities discussed in
the context of the Galilean satellites \cite[e.g.,][]{seuf11}. The magnetic field
measurements by the Galileo spacecraft near Ganymede have been
searched for signs of induction signals from an ocean by
\cite{kive02}. It was found that the magnetic field measurements
from multiple flybys are consistent with an ocean, however the
measurements can be fitted quantitatively equally well by unknown
quadrupole moments of Ganymede's dynamo magnetic field
\cite[]{kive02}. Unfortunately, it is impossible to overcome this
uncertainty, i.e., to separate spatial and temporal variability with
subsequent flybys along different trajectories. 

\begin{figure}[t]
\sidecaption
\includegraphics[scale=0.45]{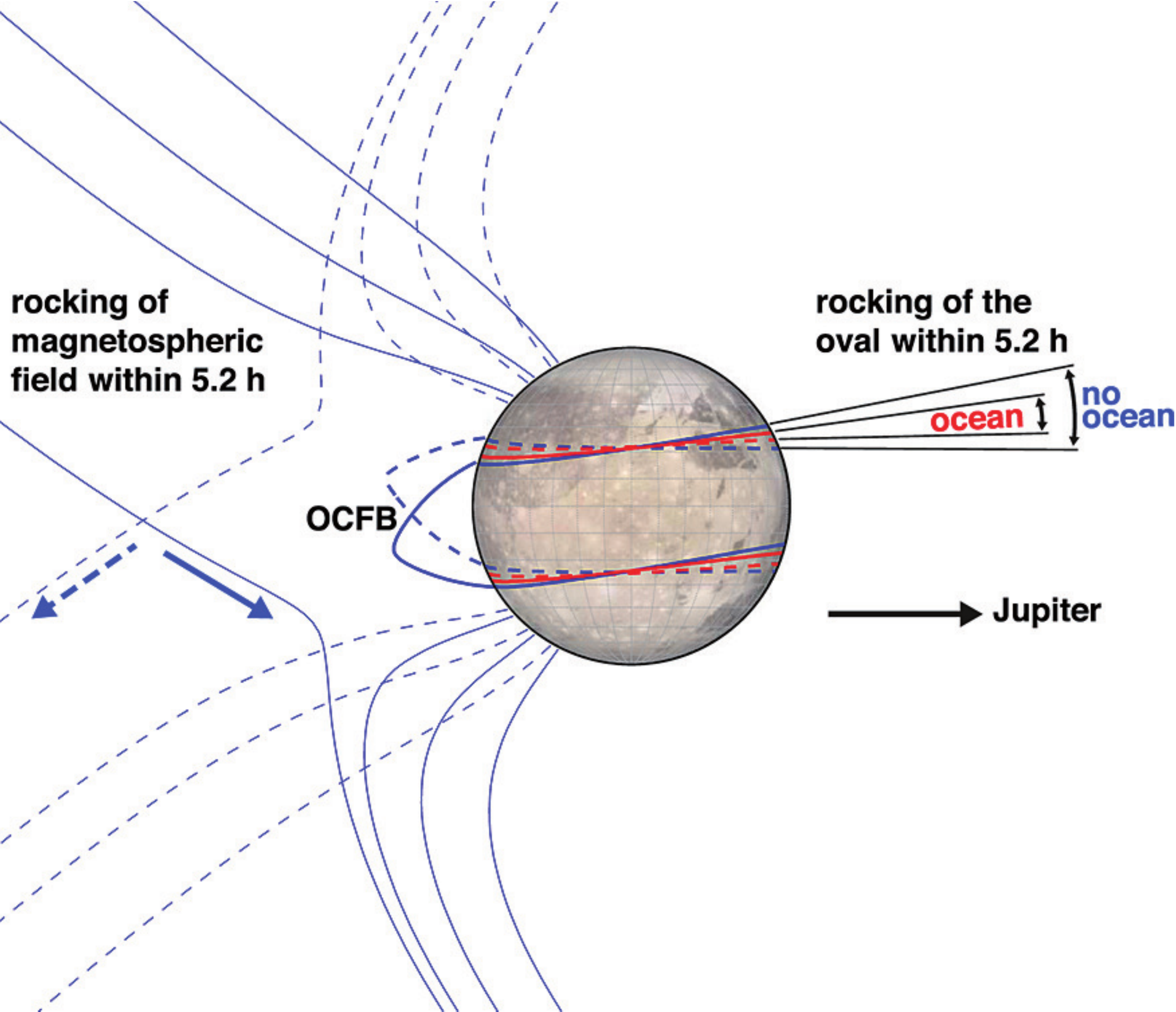}
\caption{Sketch of magnetic field lines and locations of auroral ovals when Ganymede is above (dashed lines)
and below the current sheet (solid lines), respectively. The ovals are located where the open-closed field line boundary
(OCFB) intersects Ganymede's surface. Induction in an ocean partly compensates Jupiter’s time-variable field and thus
reduces the oscillation of the ovals (red: with ocean; blue: without ocean) 
  \cite[from ][]{saur15}.
}
\label{f_sketch}   
\end{figure}

As discussed in section \ref{s_time} and visible in Figure
\ref{f_Ganymede_time_variable}, the time-variable magnetic field
component of Jupiter's magnetosphere with the
largest amplitude is the $B_r$ component 
%(i.e., radially away from Jupiter) 
with values of $\sim$80
nT. In Figure \ref{f_sketch}
we show a sketch of the magnetic field environment around Ganymede for
maximum positive $B_r$, i.e., pointing away from Jupiter (dashed
lines) and for maximum negative $B_r$, i.e., pointing towards Jupiter
(solid blue lines). The time-variable exterior component also modifies
the open-closed field line boundary (OCFL) region of Ganymede's
magnetosphere  as displayed in Figure \ref{f_sketch}) as well.  

\begin{figure}[t]
\sidecaption\includegraphics[scale=0.53]{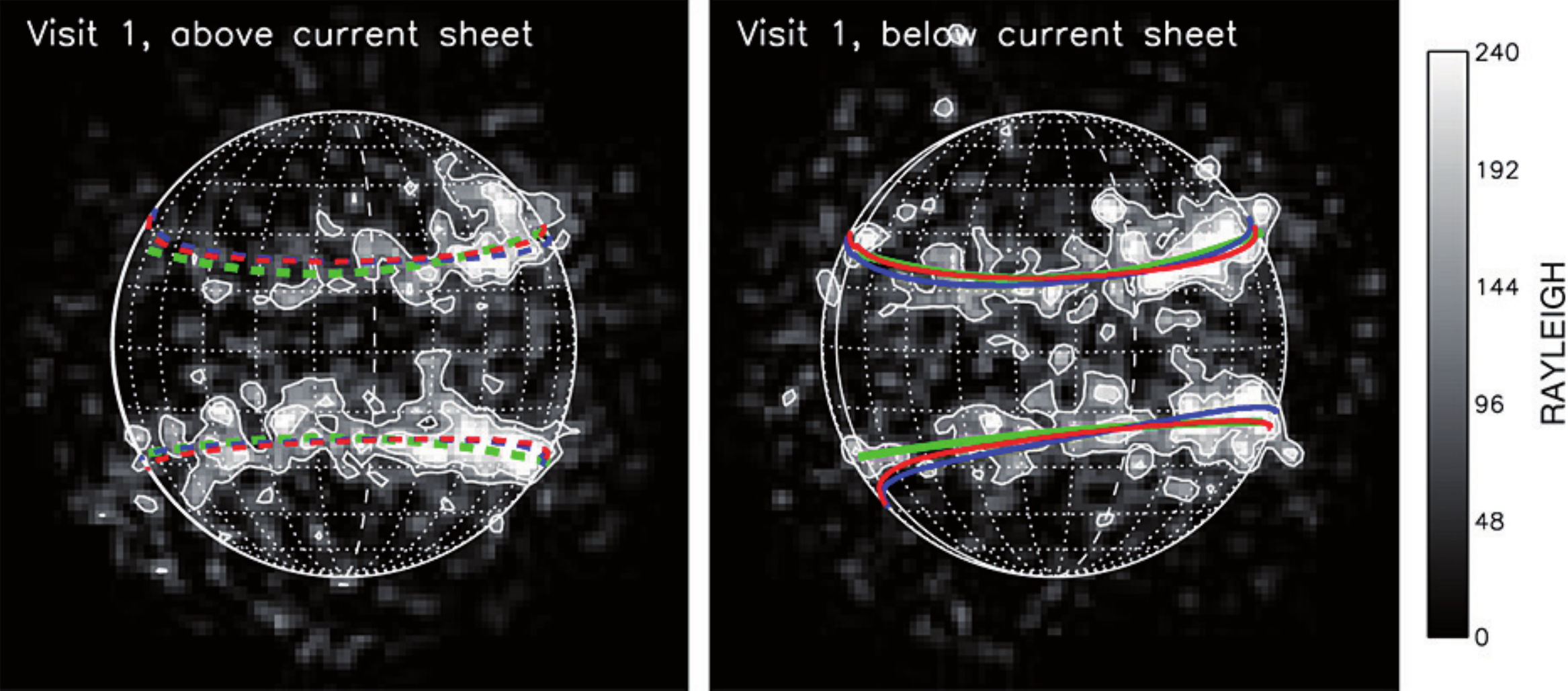}
\caption{Observed auroral brightness in Rayleigh of OI 1356 $\AA$ emission 
when Ganymede is (left) above and (right) below
the current sheet. Contours are for 110 and 170 Rayleigh. North is up and Jupiter to the right. Green lines display fits to
the observation, red and blue lines display model locations with and without ocean, respectively
  \cite[from ][]{saur15}.
}
\label{f_HST}   
\end{figure}

Ganymede also possesses two auroral ovals \cite[]{hall98,feld00,mcgr13} similar to all known planetary
bodies with an intrinsic dynamo magnetic field and an atmosphere. An
example of two auroral images taken with the Hubble Space Telescope
(HST) is shown in Figure \ref{f_HST}.
The two auroral ovals are located near the region where
the open-closed field line region intersects with Ganymede's
atmosphere as shown as red and blue lines on the surface of Ganymede
in Figure \ref{f_sketch}. Because the time-variable external magnetic
field modifies the open-closed field line boundary, the location of the
auroral ovals are time-variable as well and oscillate in the way
depicted by the solid and dashed blue lines on the disk of Ganymede
shown in Figure \ref{f_sketch}. The dashed/solid lines represents
the location of the auroral ovals when Ganymede is above/below the
%EC's comment: north/south is better (I think) than above/below
%JS's comment: that's a good point, but in the original papers we had
%used above/below. Therefore it is probably easier for the reader if
%we jumps from here to the original papers.
plasma sheet of Jupiter's magnetosphere, respectively.
When a saline and thus electrically
conductive ocean is present, the time-variable external field will
induce secondary magnetic fields which will reduce the primary
time-variable external magnetic field. The existence of an ocean will
thus also reduce the amplitude of the oscillation of the locations of
the auroral ovals. The reduced oscillation amplitudes are shown in
Figure \ref{f_sketch}, where the solid and dashed red lines indicate
the locations of the auroral ovals when an ocean is present in
contrast to the blue lines when no ocean is present. 

With dedicated HST observations obtained in November 2010 and October
2011, \cite{saur15} measured the locations of the ovals when Ganymede
is maximum above and maximum below the current sheet in the search for
a subsurface ocean. Figure \ref{f_HST} shows the HST observations
from November 2010. Figure \ref{f_HST} 
also displays  averaged locations of the ovals in green obtained from a polynomial
fit to the observed ovals. The blue and the red lines in this Figure
show the expected locations of the ovals when an ocean and no ocean are
present. The expected locations are calculated with the new MHD model of
Ganymede by \cite{duli14} described in section \ref{s_Ganymede}. Only
from visual inspections of Figure \ref{f_HST}, it is nearly impossible
to distinguish if the MHD runs with or without ocean fit better to the
observations. We note that for this purpose the absolute locations of the ovals are
not the important quantities, but the changes of the
locations, i.e., the oscillation amplitudes (see Figure \ref{f_sketch})
between the locations when Ganymede is above and below the current sheet. Therefore
\cite{saur15} calculated these difference and associated them to an
average oscillation angle, also called rocking angle $\alpha$. The average
rocking angle of the northern and southern ovals from the HST
observations in 2010 and 2011 combined was found to be
$\alpha=2.0^\circ$.

\begin{figure}[t]
\sidecaption
\includegraphics[scale=0.45]{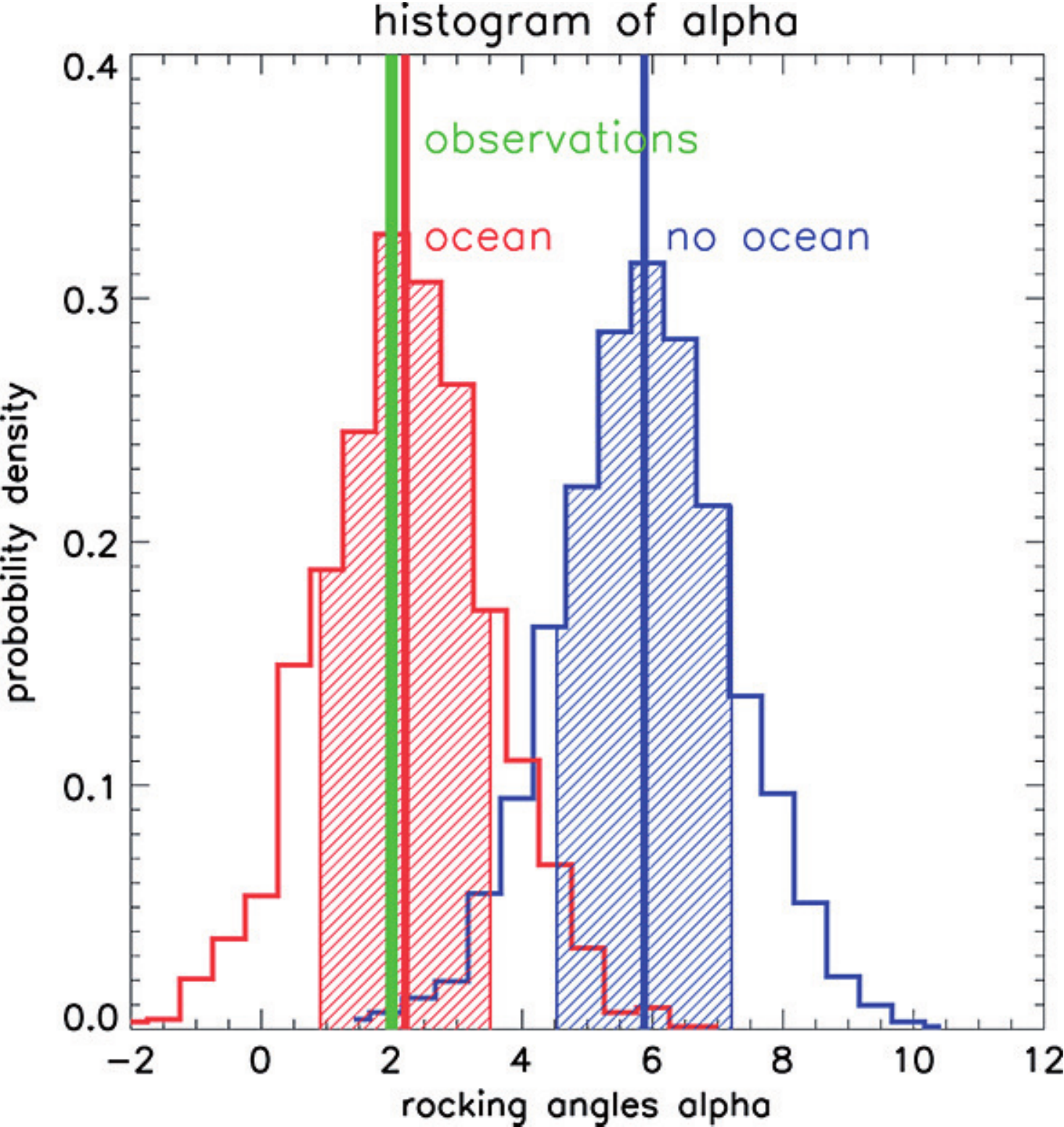}
\caption{
Distribution function of modeled rocking angles $\alpha$ from a
Monte-Carlo test with
and without ocean, respectively. The test includes the effects of stochastic
patchiness on the measurements. The vertical green line indicates
rocking angle derived from observations. The vertical red and blue
lines indicate expectation values of $\alpha$, and the 
shaded area displays the
1 sigma area around the expectation values
  \cite[from ][]{saur15}.
}
\label{f_MC}   
\end{figure}

The observed locations of Ganymede's auroral ovals in the HST data are
unfortunately patchy (see, e.g. Figure \ref{f_HST}). This patchiness
is due to intermittent reconnection near Ganymede's magnetopause and
due to the finite signal to noise ratio of the counts on the
individual detector pixels \cite[]{saur15}. In order to
assess the error when comparing model oscillation amplitudes $\alpha$
with and without ocean to the observations, \cite{saur15} introduced a
Monte-Carlo test. In the Monte-Carlo test synthetic observations with
and without ocean are generated based on the MHD model of
\cite{duli14}. Therefore patchiness produced with a random generator 
based on the physics of the intermittent
reconnection and based on the finite signal to noise of the observations
were added to the modeled locations of the ovals. The
resultant synthetic images appear visually very similar to the actually
observed ovals (see Figure 7 in \cite{saur15}). Subsequently 1024
synthetic HST campaigns where generated with individually different
patchy ovals. These synthetic images were then analyzed in an identical
ways compared to the real data. In Figure \ref{f_MC}, 
the resultant distribution of the rocking angles are shown with and
without ocean (red and blue distribution, respectively). It can be
seen that both distribution functions barely overlap, which implies
that the ocean and the non-ocean hypotheses can be well separated
with this approach. The expectation value  is 2.2$^\circ
\pm 1.3^\circ$ for the ocean model and 5.8$^\circ \pm 1.3^\circ$ for the model without
ocean. The uncertainties are calculated based on the one-sigma area
around the expectations values in Figure \ref{f_MC}. The observed
rocking angle $\alpha=2.0^\circ$ is thus consistent with the existence
of a subsurface ocean and inconsistent with no ocean present below the
surface of Ganymede. 

With this new approach the non-uniqueness of the interpretation of
the Galileo magnetometer measurements by \cite[]{kive02}
could be overcome. The key advantage of the new HST technique
by \cite{saur15} is
to search for an ocean  with time-resolved two-dimensional observations of
the auroral ovals, i.e. to apply time-dependent ''quasi
two-dimensional images'' of Ganymede's magnetic field environment.

\begin{figure}[t]
\sidecaption
\includegraphics[scale=0.4]{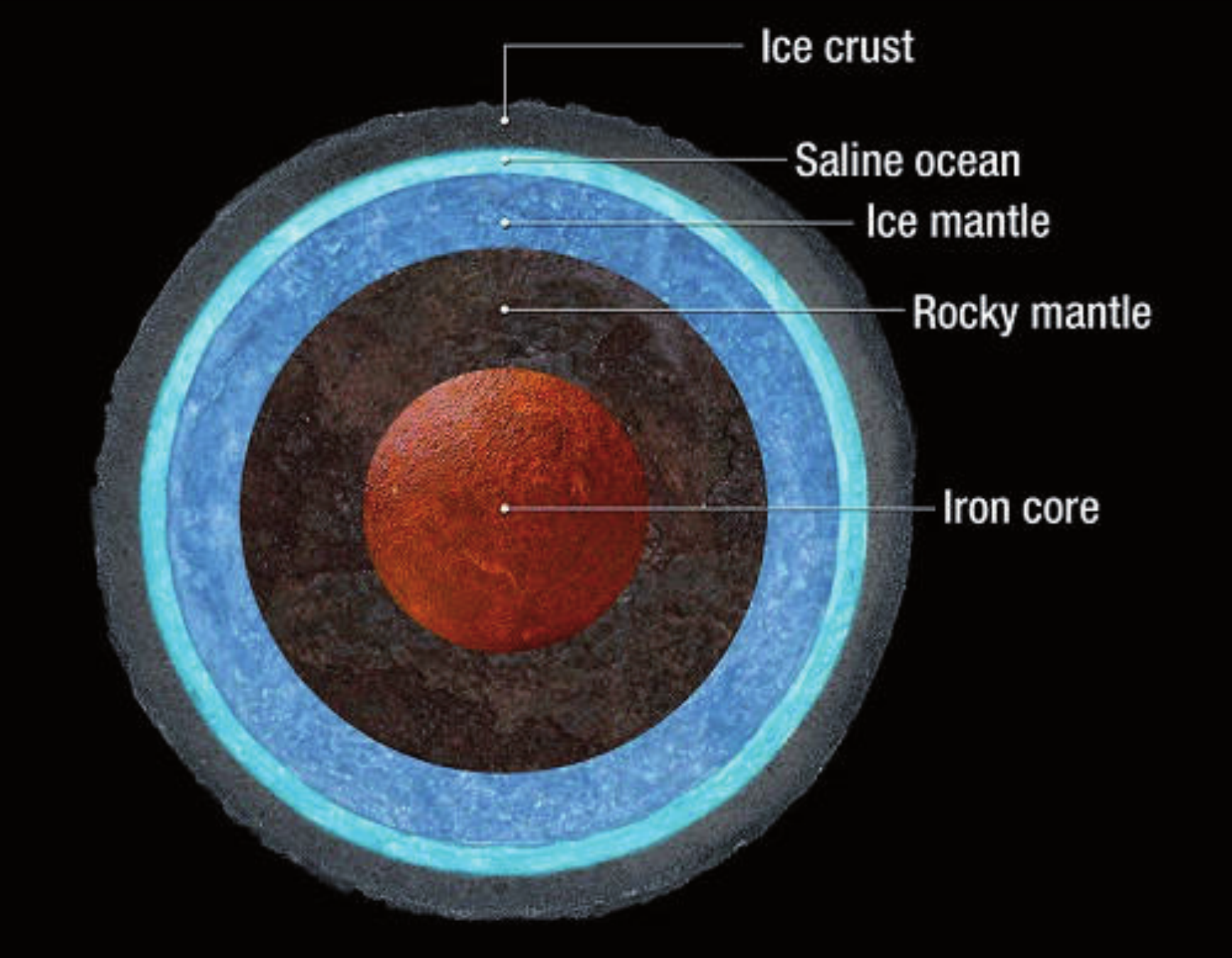}
\caption{Sketch of the internal structure of Ganymede
(Image credit NASA/STScI).
}
\label{f_interior}   
\end{figure}

The resultant internal structure of Ganymede is shown in Figure \ref{f_interior}.
The layer of liquid water is embedded within two layers of 
water in the solid phase. This structure is consistent with
theoretical models by
\cite{sohl02,huss06,ramb11,vanc14}. Based on the calculations in
\cite{saur15}, the ocean needs to have a minimum electrical conductivity
of 0.09 S/m when assuming it to be located between 150 km and 250 km
depth. This conductivity corresponds to a minimum salt concentration of
0.9 gr of MgSO$_4$ per kilogram ocean water. The measurement also
require that the top of the ocean cannot be deeper than 330 km when
measured from the surface. 

%The existence of an ocean and the presence of induced magnetic field
%signatures. 

\section{Callisto}
\label{s_Callisto}
Callisto is in size and average mass density similar to Ganymede, but
structurally only a partially differentiated body
\cite[e.g.,][]{show99}. Callisto
 also does not possess an intrinsic dynamo magnetic field in contrast to
 Ganymede. However, it encompasses similar to the other Galilean
 satellites a thin atmosphere and an ionosphere
 \cite[]{carl99,klio02,cunn15}.
%In contrast to Ganymede, Callisto does not possess an intrinsic dynamo magnetic field, but the time-variable component of Jupiter's magnetospheric field similarly induces also within a subsurface ocean within Callisto secondary magnetic fields \cite[]{zimm00}. 
The time-variable components of Jupiter's magnetospheric field 
induce electric currents within electrically conductive layers, such
as a subsurface ocean, creating an induced dipole magnetic field based on studies by
\cite[]{neub98} and \cite{zimm00}. 
Callisto's atmosphere and ionosphere interact with the plasma of Jupiter's magnetosphere, which generates additional magnetic field perturbations.
In the following two subsections the formation of Callisto's
ionosphere and the plasma interaction will be discussed.

\subsection{Callisto's ionosphere}
 The first component of Callisto's
 atmosphere to be observed was CO$_2$ with a column density of 0.8
 $\times$ 10$^{19}$ m$^{-2} $ \cite[]{carl99}. The large ionospheric
 densities up to 4 $\times 10^{10}$ m$^{-3}$ inferred by \cite{klio02}
 imply that an additional atmospheric component is present, which
 was suggested to be O$_2$. Hubble Space Telescope observations
 with the STIS camera however only led to upper limits for O$_2$
 \cite[]{stro02}. But subsequent observations presented by \cite{cunn15} with the more
 sensitive HST/COS camera revealed OI 135.6 nm and 130.4 nm
 emission with a  brightness of 1 to 5 Rayleigh. The authors derived from these 
 observations an O$_2$ column density of 4 $\times$ 10$^{19}$ m$^{-2}
 $.

\begin{figure}[t]
\sidecaption
\includegraphics[scale=0.8]{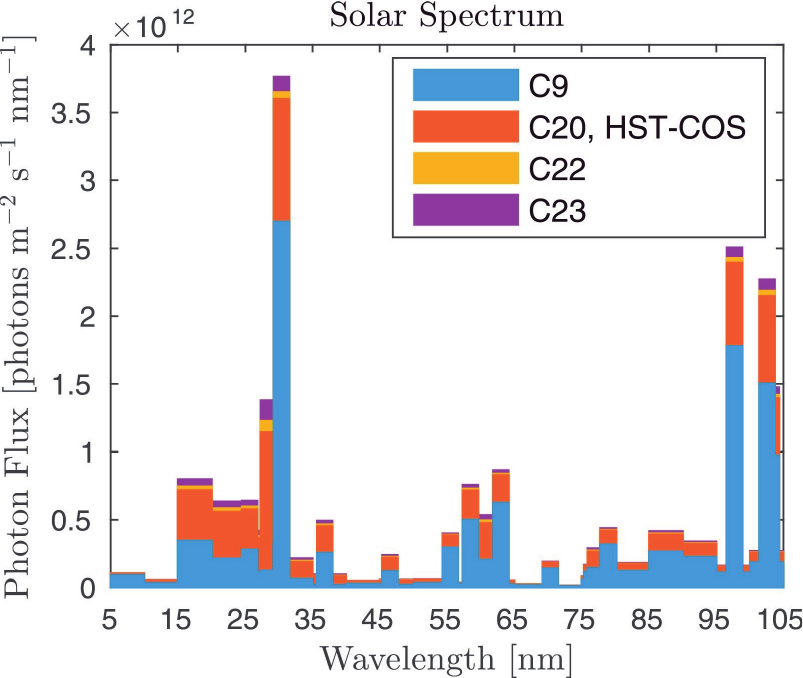}
\caption{Variability of the 
solar photon fluxes at Callisto's solar distance for the times of the
Callisto C-9, C-20, C-22, C- 23 flybys and the HST/COS
observation. The fluxes are plotted overlapping. The largest fluxes
occurred during C-23 while the smallest fluxes occurred during
C-9. The according dates are: C-9: 1997/06/25, C-20: 1999/05/05, C-22:
1999/08/14, C-23: 1999/09/16, HST/COS: 2011/11/17
\cite[from][]{hart17}.
}
\label{f_solar}   
\end{figure}

A main difference of the ionosphere and the atmospheric
UV emission of Callisto compared to those of the other Galilean
satellites is that the plasma density of Jupiter's magnetosphere at
Callisto is so dilute that electron impact is not the primary source
of ionization and UV excitation anymore. Callisto's ionosphere and UV
emission is in contrast primarily driven by solar photons
\cite[]{cunn15}. In order to better understand Callisto's atmosphere and
ionosphere, \cite{hart17} developed a new model to simultaneously
explain the observed ionospheric electron column densities and the
atmospheric UV emissions. This model solves for the electron
distribution functions at every location in Callisto's atmosphere for a
prescribed atmosphere which includes O$_2$, CO$_2$, and H$_2$O. It
takes into account as the primary source of electrons 
the solar UV fluxes which are highly time-variable
as displayed in Figure \ref{f_solar}.
The model solves the Boltzmann equation for the supra-thermal electron
population and considers a large set of inelastic collisions between
the atmospheric species, which modifies the
electron energies. As a loss for electrons recombination is included,
which is energy dependent and more effective for lower electron
energies. The model neglects spatial transport of the electrons, which
is a good assumption for altitudes smaller than 180 km for the thermal
electrons and smaller than 45 km for supra-thermal electrons \cite[]{hart17}. For the low
temperature electrons, i.e., energies approximately less than 0.5 eV,
the electron distribution function is Maxwellian due to the importance of 
electron-electron collisions at these energy ranges. Therefore the
model of \cite{hart17} describes the electrons in this energy range, referred
to as thermal electron range, with a fluid description for the electron
particle densities and energy densities. 

\begin{figure}[t]
\sidecaption
\includegraphics[scale=0.8]{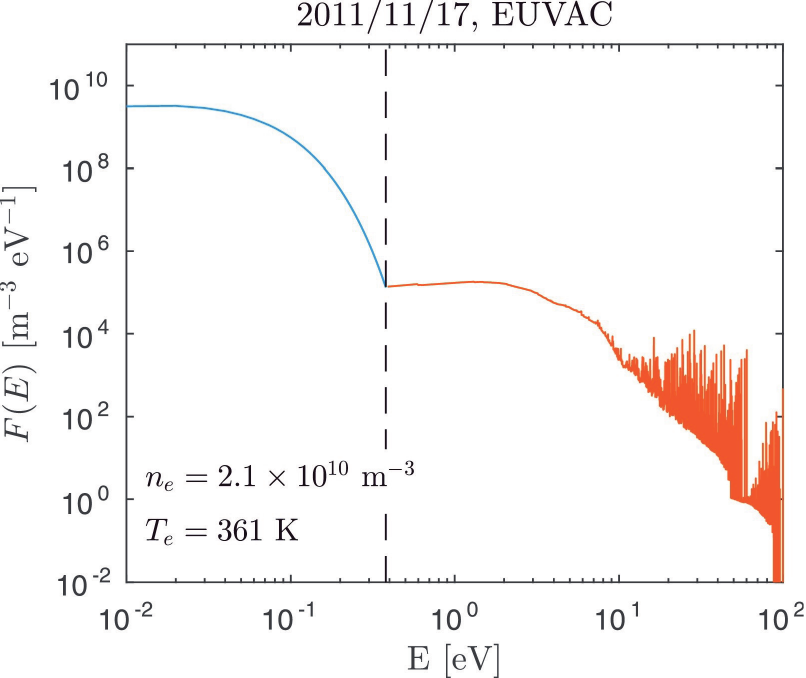}
\caption{
Calculated electron energy distribution function of a volume element
in Callisto's ionosphere.
The prescribed neutral densities are $1.0 \times
10^{15}$ m$^{-3}$ for O$_2$  and 
$0.33 \times 10^{15}$ m$^{-3}$ for CO$_2$.   
Resulting electron density and temperature are $2.1 \times
10^{10}$ m$^{ -3}$ and $T_e$ = 361 K.
The dashed black line marks the transition from the kinetic to the
fluid range, which is located for this volume element at 0.38 eV
\cite[from][]{hart17}.
}
\label{f_electron_dist}   
\end{figure}

A resultant electron distribution function from the model of
\cite{hart17}
is shown in Figure \ref{f_electron_dist}.
It demonstrates the highly non Maxwellian nature of the distribution
function for energies larger than 0.5 eV. This distribution function
at these energies can only be calculated with a kinetic model. It shows the imprints of
the solar input spectrum and the large class of possible collisional processes
with the atmospheric neutrals. At lower temperature the
distribution function turns Maxwellian due to the electron-electron
collisions.
 
For a joint interpretation of the observed UV emission and the
ionospheric electron densities a kinetic description of the electrons is
necessary. The high energy tail of this distribution function excites the UV
emission emitted from Callisto's atmosphere. With a plain Maxwellian
distribution with the temperature of the thermal population
barely any UV radiation would be emitted.  

\begin{figure}[t]
\sidecaption
\includegraphics[scale=0.8]{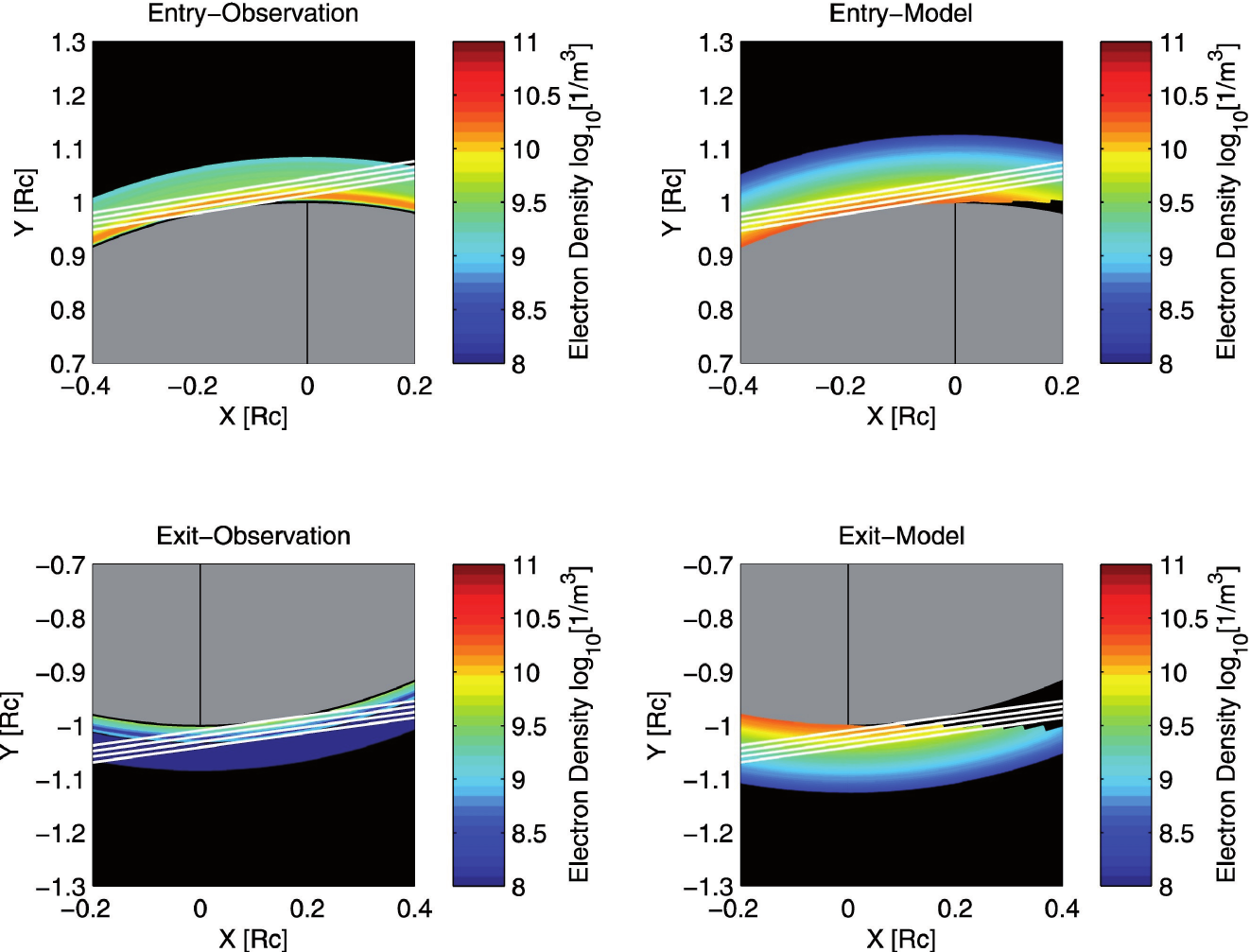}
\caption{
Electron densities of the terminator regions in the equatorial plane
according to the C-22 entry and exit electron density altitude
profiles of \cite{klio02} (left panels) and according to an
exemplary model ionosphere with configurations of C-22 (right
panels). For the shown model results (right panels), the prescribed
atmosphere is spherically symmetric with an O$_2$ column density of 
$3.0 \times ×10^{19}$ m$^{ -2}$. White lines correspond to radio occultation LOS during
entry (upper row) and exit (lower row) of flyby C-22. In this
Cartesian coordinate system, the Sun is in the -x direction and the
y-axis is in the equatorial plane. Length scales are given in units of
Callisto’s radius R$_C$ = 2410 km
\cite[from][]{hart17}.
}
\label{f_electron_iono}   
\end{figure}

With the combined kinetic and fluid model, \cite{hart17} calculate
electron densities in Callisto's ionosphere. Several examples of the
resultant electron density structure are displayed in Figure \ref{f_electron_iono}.
The figure also shows line of sight paths of the Galileo spacecraft
radio science signals through which ionospheric electron densities
were derived by \cite{klio02}. The radio science observations were
all taken around the terminator region and thus provide constraints on
Callisto's ionosphere only in this region. Because the radio science
technique constrains the integral electron density along a line of
sight only, this effect has to be considered when comparing
model results with observations. 

With the kinetic-fluid model of \cite{hart17}
the HST observations of \cite{cunn15} and the
electron density measurement of \cite{klio02} 
can be jointly explained within a certain
range of O$_2$ densities in Callisto's atmosphere. Based on this
comparison, Callisto’s atmosphere has a mean O$_2$ column density of
$2.1  \pm 1.1  \times 10^{ 19}$ m$^{ −2}$ and 
the atmosphere possesses a day night
asymmetry. The terminator O$_2$ column density has values of 
$\sim0.4  \times 10^{ 19}$ m$^{ -2}$ and associated subsolar O$_ 2$
column densities are in the range of
$2.4 - 9.8  \times 10^{ 19}$ m$^{ -2}$ . The calculations by
\cite{hart17}
also show that the electron
density is very sensitive to the relative abundance of H$_2$O in
Callisto's atmosphere due to
the thermal electron cooling by rotational state excitation of
H$_2$O. For the efficiency of Callisto’s atmospheric UV emission it is found
that on average one photon is emitted at OI 135.6 nm per every 170
electron ion pairs generated and per every 60 electron ion pairs
produced by secondary electron impact ionization.

\subsection{Callisto's plasma interaction}

Callisto with its atmosphere and ionosphere is an obstacle to the
dilute plasma of Jupiter's magnetosphere streaming past the moon. This
interaction is characterized by  a magnetospheric 
space plasma environment which is more variable compared to the other Galilean
satellites. The magnetospheric field $B_0 = 4-42$ nT, the relative
velocity $v_0=122-272$ km/s, the ion density $n_0= 0.01-0.5 \times
10^6$ m$^{-3}$, the Alfv\'en Mach numbers 
M$_A=0.02-1.85$ and the ion gyro radius $r_g$ = 34-530 km are highly
variable due to the varying  position of Callisto with respect to
Jupiter's magnetospheric plasma sheet and due to stochastic effects within the
magnetosphere \cite[]{kive04,seuf12}.

\begin{figure}[t]
\sidecaption
\hspace*{-2.cm}
\includegraphics[scale=0.6]{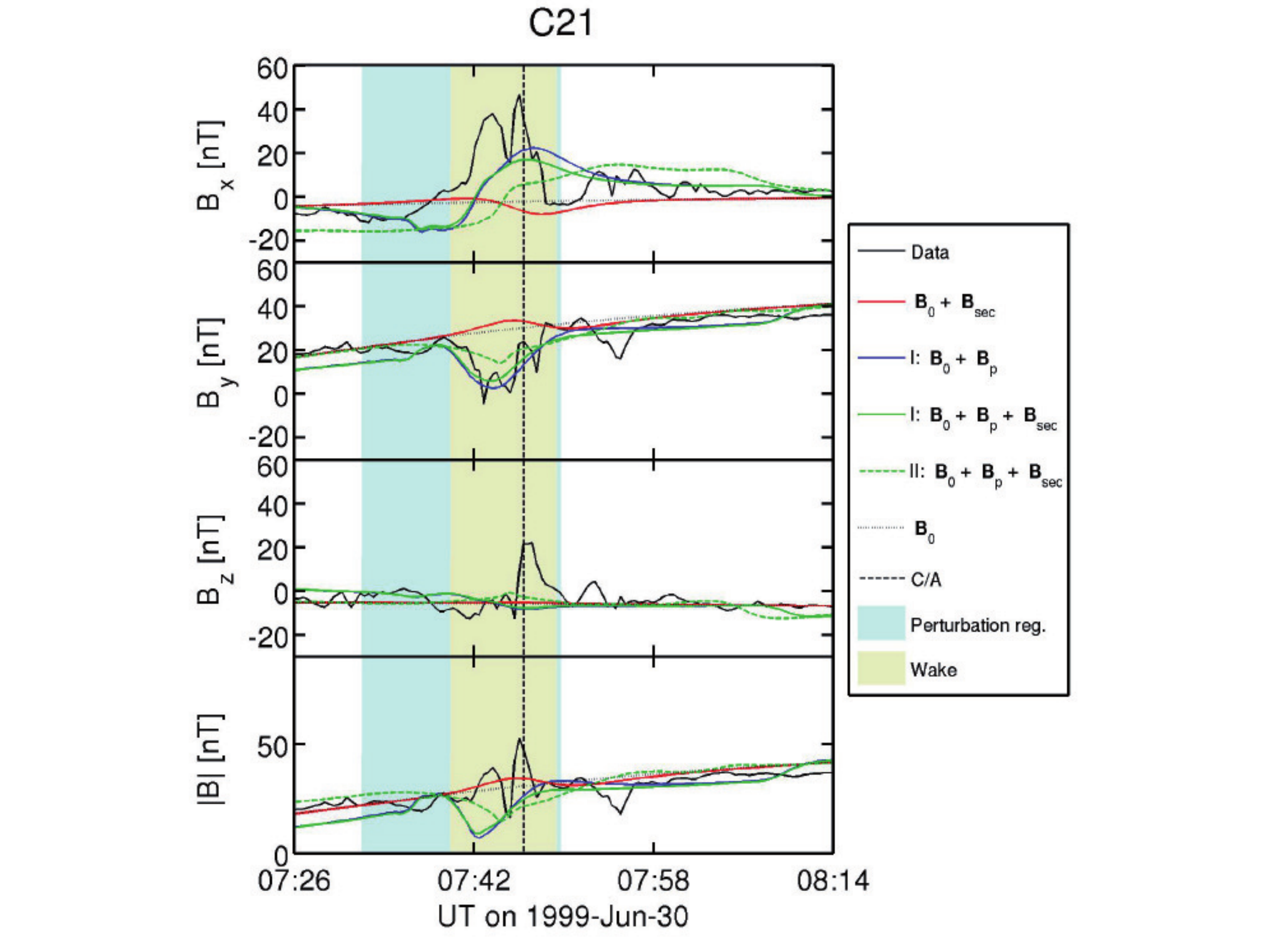}
\caption{
%Magnetic field measurements and modeled magnetic fields along
 % the C-21 flyby trajectory of the Galileo spacecraft. 
%{\bf 
Magnetic field measurements along the C-21 fly trajectory of the
  Galileo spacecraft (black solid lines) in comparison with modeled
  magnetic fields in nT. 
The superposition of the background magnetic field (black dotted
lines)  and the induced
fields is shown in red. 
The blue solid lines depict a superposition of the background field and the
modeled plasma interaction fields for a model (case I)
 using the measured plasma data with a velocity
artificially decreased by a factor of five. The green solid lines represent a 
superposition of the plasma
interaction and the induced fields for the case I as well.  
The green dashed 
lines give a similar superposition
for the default model in \cite{seuf12}, i.e., using a corotational
plasma velocity of 192 km/s (case II). 
Blue and ocher areas indicate the locations where the
perturbation region and the geometrical wake for case I 
are crossed by the spacecraft trajectory. 
The vertical dashed line indicates the time of the closest approach
%}
\cite[from][]{seuf12}.
}
\label{f_C21}   
\end{figure}

\cite{seuf12} constructed an MHD model of Callisto's interaction with
Jupiter's magnetosphere similar to the model for Ganymede described in
section \ref{s_Ganymede}. The model includes the formation of an
ionosphere through photoionization and electron impact ionization
within a pure CO$_2$ atmosphere and alternatively within an atmosphere
composed of CO$_2$ and O$_2$. The model
also includes as internal magnetic fields the induced fields from a
subsurface ocean within Callisto (see section \ref{s_time}). The
resultant magnetic fields from the MHD model of \cite{seuf12} in
comparison with measurements made by the Galileo spacecraft during
the C21 flyby are displayed in 
Figure \ref{f_C21}.
The C-21 flyby was a flyby through the wake of Callisto with a closest
approach of about 1000 km. In Figure \ref{f_C21}, the observed fields
are shown in black and the induced magnetic field combined with the
background field are
shown in red. The magnetic field including the plasma interaction with
with a CO$_2$ atmosphere only is shown as  green dashed line  
and the field from the plasma interaction  with a combination of CO$_2$ and
O$_2$ is shown as  solid green line. 
The results demonstrate that both  induction in a conductive ocean  and the
plasma interaction produce significant magnetic field
perturbations. Similar conclusions have been reached by subsequent
modeling of \cite{liuz15,liuz16}. Therefore an
appropriate modeling of the plasma interaction and the induction
effects  is necessary in the
interpretation of the plasma and field measurements obtained by the Galileo
spacecraft. 

\section{Summary}
In this chapter, we discussed similarities and differences between the
largest and one of the smallest magnetosphere in the solar systems,
i.e., those of Jupiter and Ganymede. We introduced two new models for
the description of their magnetic field and plasma
environments. These models cover two classes of the plasma
interaction between a magnetized body with its surrounding space
plasma, i.e., sub-Alf\'enic and super-fast interactions, which 
have counterparts at extrasolar planets. We also showed how 
HST observations in conjunction with MHD
modeling of Ganymede's auroral
ovals can be used to search for a subsurface ocean within Ganymede. We
additionally investigated the
non-Maxwellian nature of the electron distribution function in Callisto's
ionosphere to constrain its atmosphere based on HST and Galileo
spacecraft measurements.

\input{reference_saur~}
\end{document}